\DocumentMetadata{}
\documentclass[sigconf]{acmart} 

\AtBeginDocument{%
  \providecommand\BibTeX{{%
    \normalfont B\kern-0.5em{\scshape i\kern-0.25em b}\kern-0.8em\TeX}}}

\copyrightyear{2025}
\acmYear{2025}
\setcopyright{cc}
\setcctype{by}
\acmConference[C\&C '25]{Creativity and Cognition}{June 23--25, 2025}{Virtual, United Kingdom}
\acmBooktitle{Creativity and Cognition (C\&C '25), June 23--25, 2025, Virtual, United Kingdom}\acmDOI{10.1145/3698061.3726929}
\acmISBN{979-8-4007-1289-0/2025/06}

\makeatletter
\def\@ACM@copyright@check@cc{}
\makeatother
\setcopyright{cc}

\usepackage{stfloats}
\usepackage{graphicx}
\usepackage{subcaption}
\usepackage{soul}
\usepackage{comment}
\usepackage{tikz}
\usetikzlibrary{shapes.geometric, arrows}
\usepackage{hyperref}
\usepackage{hyperxmp}
\usepackage{array,multirow}
\usepackage{draftwatermark}

\begin{document}

\title[Kaleidoscope Gallery]{Kaleidoscope Gallery: Exploring Ethics and Generative AI Through Art}

\author{Alayt Issak}
\authornote{Corresponding author issak.a@northeastern.edu.}
\email{issak.a@northeastern.edu}
\orcid{0009-0005-8149-7358}
\affiliation{%
  \institution{Northeastern University}
  \streetaddress{360 Huntington Avenue}
  \city{Boston}
  \state{MA}
  \country{USA}
  \postcode{02115}
}
\renewcommand{\shortauthors}{Issak et al.}

\author{Uttkarsh Narayan}
\email{narayan.u@northeastern.edu}
\affiliation{%
  \institution{Northeastern University}
  \streetaddress{360 Huntington Avenue}
  \city{Boston}
  \state{MA}
  \country{USA}
  \postcode{02115}
}

\author{Ramya Srinivasan}
\email{ramya@fujitsu.com}
\affiliation{%
  \institution{Fujitsu Research of America}
  \streetaddress{4655 Great America Pkwy, Suite 410}
  \city{Santa Clara}
  \state{CA}
  \country{USA}
  \postcode{95054}
  }

\author{Erica Kleinman}
\email{e.kleinman@northeastern.edu}
\affiliation{%
  \institution{Northeastern University}
  \streetaddress{360 Huntington Avenue}
  \city{Boston}
  \state{MA}
  \country{USA}
  \postcode{02115}
}

\author{Casper Harteveld}
\email{c.harteveld@northeastern.edu}
\affiliation{%
  \institution{Northeastern University}
  \streetaddress{360 Huntington Avenue}
  \city{Boston}
  \state{MA}
  \country{USA}
  \postcode{02115}
}

\renewcommand{\shortauthors}{Issak et al.}

\begin{abstract} 
   Ethical theories and Generative AI (GenAI) models are dynamic concepts subject to continuous evolution. This paper investigates the visualization of ethics through a subset of GenAI models. We expand on the emerging field of Visual Ethics, using art as a form of critical inquiry and the metaphor of a kaleidoscope to invoke moral imagination. Through formative interviews with 10 ethics experts, we first establish a foundation of ethical theories. Our analysis reveals five families of ethical theories, which we then transform into images using the text-to-image (T2I) GenAI model. The resulting imagery, curated as Kaleidoscope Gallery and evaluated by the same experts, revealed eight themes that highlight how morality, society, and learned associations are central to ethical theories. We discuss implications for critically examining T2I models and present cautions and considerations. This work contributes to examining ethical theories as foundational knowledge that interrogates GenAI models as socio-technical systems.
\end{abstract}



\maketitle
\SetWatermarkText{PRE-PRINT}
\SetWatermarkScale{1}

\section{Introduction}
\label{intro}

Ethics involves understanding what is right and wrong in relation to the world and how to deal with uncertainty~\cite{moran_introduction_2014}. Ethical theories are the formalizations of guidelines to inform ethical decision-making. The theories change as they are situated in definition and praxis as a result of changes to statutes and practices, respectively~\cite{theory_and_practice}. Technology---the application of knowledge to produce new capabilities---has historically been used to critically examine ethics. This has occurred indirectly through innovations that prompt ethical dilemmas~\cite{scott_surrogacy_2009, landecker_dna_1999, bennett_agricultural_2013}, or directly with interdisciplinary efforts such as critical AI Art~\cite{walker_ai_2023, whitney_critical_2023}. Present day, technology serves as both a catalyst and a tool for ethical reflection. 

In this work, we focus on GenAI models as technological innovations for critically examining how ethics is represented through these technologies. Similar to ethics, GenAI models change as they rely upon model training that invites continuous change~\cite{wang_watch_2023}. This continuous change, as a result of human beings' interaction with GenAI models, has recently been proposed as Human-AI coevolution~\cite{co-evolution}. Widespread mechanisms of model training, or how Human-AI coevolution occurs, include alignment to human preferences in model learning~\cite{lee_rlaif_2023} and unlearning~\cite{yao_unlearning_2023}, fluid data that ascribes itself~\cite{offenhuber_autographic_2023}, training of GenAI models on synthetic data (a method that has been criticized for model collapse when used recursively)~\cite{shumailov_ai_2024,alemohammad_self-consuming_2023}, and prompt tuning from real-world user input~\cite{openai_train_2024}. Deploying GenAI models for knowledge representation carries significant risks, including bias~\cite{naik_social_2023}, misinformed output~\cite{bird_typology_2023}, and uncertain reliability~\cite{wolfe_implications_2024}. With this in mind, our work critically examines a subset of GenAI models known as text-to-image (T2I) models~\cite{hf_t2i_2024} by using them to represent ethical concepts. T2I models convert natural language input into visual imagery, making them prime candidates for knowledge representation in GenAI models. 

For this critical examination of how T2I models represent ethics, our research employs art. Art is not merely a representation but rather a catapult of imagination, i.e., opening up new perspectives of literacy~\cite{art_literacy} and understanding~\cite{daniele_ai_2019}. It has been highlighted as a tool to investigate AI literacy as it illuminates a deeper understanding of AI systems through critical investigation, novel perspectives, and edifying experiences to audiences~\cite{hemment_ai_2023}. Previous work has highlighted the ability of the arts to foster empathy~\cite{rusu_art_empathy} and expression~\cite{art_expression} for comprehension and understanding. Our work specifically aligns with a critical examination of ethics through the emerging interdisciplinary field of \textit{Visual Ethics}. Moral Imagination, a seminal book in this field, states the essential role imagination plays in ethical deliberation~\cite{johnson_moralimagination_1994}. Art and visual display bring forth aesthetic experiences that question what is seen and what is not~\cite{srinivasan_role_2021}, such as metaphors that structure understandings of our cognition and experiences~\cite{johnson_metaphors_2003}. In relation to our work,~\citet{srinivasan_building_2021} discuss the visualization of ethical viewpoints as a component of generative artworks for the design of ethical AI systems. A similar study also suggests image generation based on ethical viewpoints to mitigate algorithmic representations~\cite{srinivasan_patent}. These relationships provide insight into visual images as a reflection of a model's design~\cite{qadri_ais_2023, cetinic_understanding_2021}. As part of our visual inquiry, we ask the following research questions: 

\begin{enumerate}
    \item How do T2I models visually represent ethical theories?
    \item How do T2I models conceptually represent ethical theories?
   \item How do experts in ethics assess the generated images to reflect their understanding of ethics? 
\end{enumerate}  

\begin{figure}[h]
    \centering
    \includegraphics[alt={Interior components of a kaleidoscope with the relationship to Ethics and GenAI}, width=\linewidth]{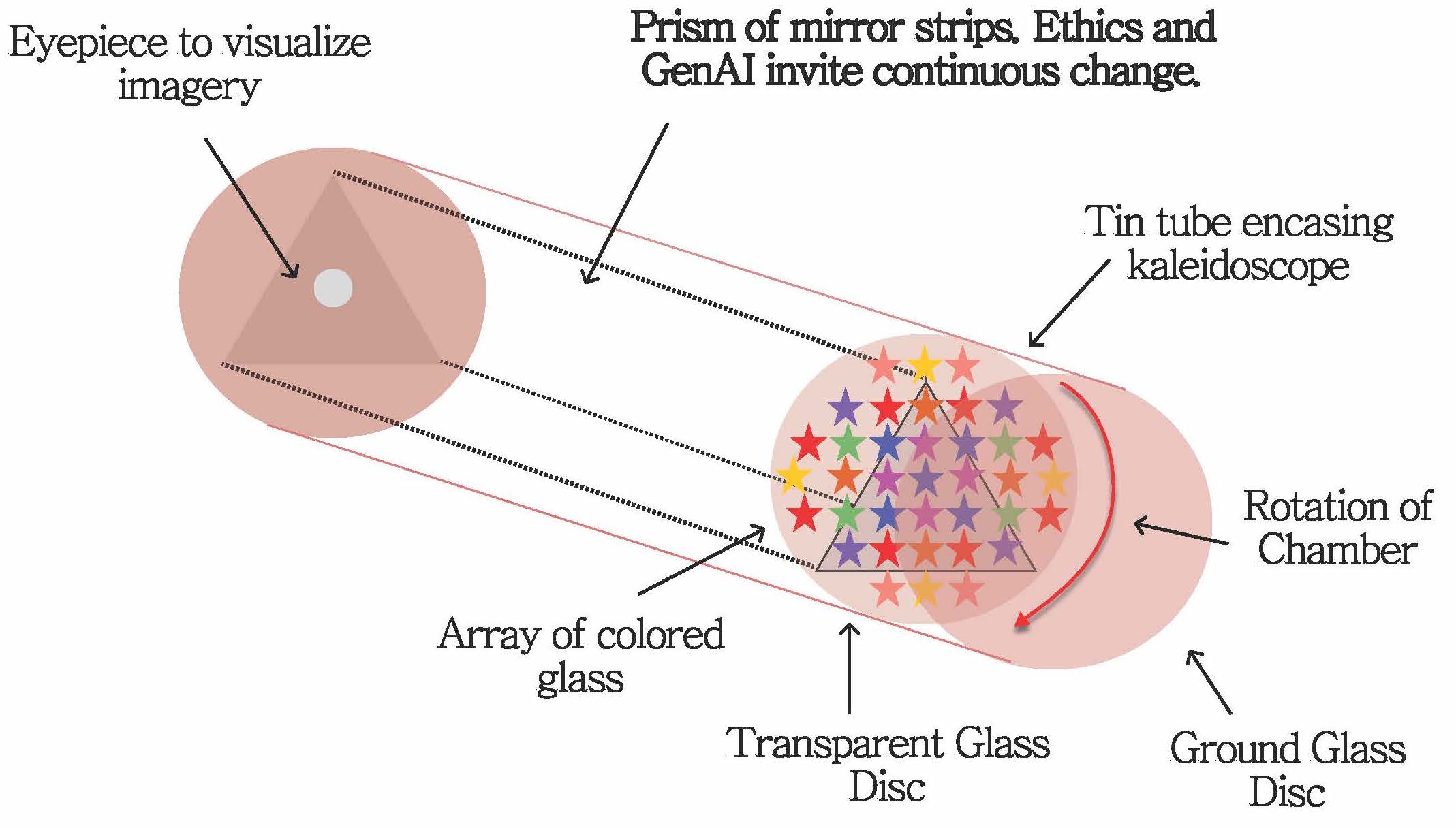}
    \caption{Mechanics of a kaleidoscope as it relates to our study (in bold text).}
    \label{fig:design}
\end{figure}

To answer these questions, we first examine ethical theories and GenAI models through the metaphor of a kaleidoscope. The kaleidoscope has been deployed as a metaphor for iterative feedback~\cite{sterman_kaleidoscope_2023} and value pluralism~\cite{sorensen_value_2023} in other works. In our work, we find the constant shifting of colors, visuals, and shapes within a kaleidoscope to encompass the dynamic state of interaction in ethics and GenAI models. We use the metaphor as it is a tool for cognition that is pervasive in everyday life~\cite{johnson_metaphors_2003}. It facilitates communication within specific conceptual systems and is a creative tool for language~\cite{flis_conceptual_metaphor_2024}. As the chamber of the kaleidoscope turns, the prism of mirrors reflects onto each other, creating a continually changing pattern of shapes and colors that alludes to the complex dynamism of GenAI and ethics (Figure \ref{fig:design}). 

Our procedure begins with a formative study among ten experts in ethics to collect viewpoints on ethical theories. We then synthesize our imagery using 
a T2I model known as DALL-E 3~\cite{betker_improving_dalle3}. We then put together a curation titled \textit{Kaleidoscope Gallery}. This gallery is used to see how ethical theories are visually represented (RQ1) and conceptualized (RQ2). Finally, we showcase the gallery to our experts and follow up with an evaluative study (RQ3). Our results showcase the diversity and fluidity of ethical theories, and the impact of worldview expression on model behavior, such as western, hierarchical and geographic bias. Overall, we aim to contribute this work to the ongoing dialogue about the role of GenAI in shaping ethical discourse and its potential to enhance our understanding of complex philosophical concepts. Specifically, we use the metaphor of the kaleidoscope to introduce GenAI models as ever-changing technologies, and that, when encountered in the real world, shape public understanding of concepts such as ethics.
This work views creativity as a driver of positive change for the societal implications of GenAI models.

\section{Background}

\subsection{Ethical Theories}

\subsubsection{Integrating Human Values in AI} 
In our work, we investigate how T2I models represent the human-centered value of ethics. Various human-centered values have been proposed for the integration of ethical foundations in AI models. These include, but are not limited to, honor~\cite{wu_honor_2023}, respect~\cite{seymour_respect_2022}, human rights~\cite{human_rights_in_AI}, and democratic rights~\cite{democratic_embedding}. Previous work has proposed an ETHICS benchmark taking a literal viewpoint toward the evaluation of models against ethical value judgments, such as justice for aligning AI with shared human values~\cite{ethics_dataset_2023}. Recent work has elaborated this integration with a pluralistic lens, in acknowledging that various human values, rights, and duties can lead to equally discernible decisions. Work by~\citet{sorensen_value_2023} highlights this pluralistic viewpoint with a kaleidoscope metaphor for the multiple value variations that exist in a model and are emitted toward a user. By utilizing the same metaphor, our work builds on the pluralistic capabilities of representing multiple theories through visualization. This visual approach extends understanding across text to image modalities which is central to our study (RQ1). 

\subsubsection{Conceptual Investigation of Artifacts} 
Our work investigates whether generative artifacts encapsulate conceptual underpinnings, and thus whether artifacts are created with the model's understanding of its generation. Multi-modal across text and image, recent work has critically detailed that a model's generative capability may not be contingent upon its understanding capability~\cite{west2023generative_understand}. To surmount this within ethics, the Delphi experiment, operationalized on prominent ethical theories, elaborates on whether machines can reason about ethical perspectives~\cite{jiang_delphi}. Limited to descriptive ethics, the Delphi experiment recognizes the need to include top-down constraints for coherent knowledge integration of machine predictions. As a mitigation to this phenomenon, work on the ``Ethics of AI Ethics" highlights conceptual, substantive, and procedural approaches to the critical analysis of the AI lifecycle and output, i.e., generation and understanding, respectively~\cite{heilinger_ethics_of_ethics_2022}. 
This leads us to inquire upon the conceptual representation of ethical theories through GenAI (RQ2). 

\subsection{GenAI models}

\subsubsection{Embedding Semantics} 
Embedding semantics in imagery is a technique that is pivotal to our work as ethics is largely understood via signs and symbols. This includes images that carry moral meaning such as oaths, scales, and gavels. Prominent examples of embedding semantics in images include the hybrid representations of text and image in pictograms, ideograms, and logograms~\cite{ernest_pictograms_2018}. Previous work by~\citet{tendulkar_trick_2019} addresses this representation for generating conceptual text via a semantic reinforcement approach known as TReAT---Thematic Reinforcement for Artistic Typography. Given an input word and a theme, TReAT replaces the input word with a clipart of each letter upon the context or theme of the word. In a similar approach, recent work by~\citet{iluz_word-as-image_2023} furthers this approach by representing textual concepts visually with the use of large language models. This approach elucidates a word-as-image for semantic typography, utilizing a pre-trained Stable Diffusion model to guide the generation of the illustration. This work extends previous methods' replacement of letters with existing icons~\cite{tendulkar_trick_2019,zhang_ornamental_2017}. In relation to our work, the incorporation of generated semantics in artifact design aligns with our representation of conceptual imagery through GenAI (RQ2). 

\subsubsection{Text-to-image (T2I) Models} 
Image generation from textual input is a process articulated by T2I models, with current models including DALL-E, MidJourney, Stable Diffusion, and Imagen, among others~\cite{T2I_survey}. Situated in the emerging field of visual ethics, we deploy these models to visualize ethical theories and invoke moral imagination. An evaluation of deep generative models, encompassing Variational Autoencoders (VAEs), Generative Adversarial Networks (GANs), and Diffusion models, reveals that each model optimizes different attributes~\cite{gainetdinov_diffusion_2023}. Specifically, VAEs excel in fast sampling and model coverage (diversity), GANs are renowned for fast sampling and producing high-quality images, while Diffusion models are noted for generating high-quality images and achieving model coverage (diversity). Additionally, the transformer architecture~\cite{wang_transform_2024} and its integration with diffusion models~\cite{diffusionmodelstransformers} are widely used. T2I models are recognized as a subset of foundation models, recently introduced as Media Foundation Models~\cite{MovieGen}, making them ideal for interrogating the representation of complex concepts such as ethics.

\section{Methodology}
\label{methods}

\begin{figure*}[!htb]
    \centering
    \includegraphics[alt={Summary of research methods as flow chart}, width=\linewidth]{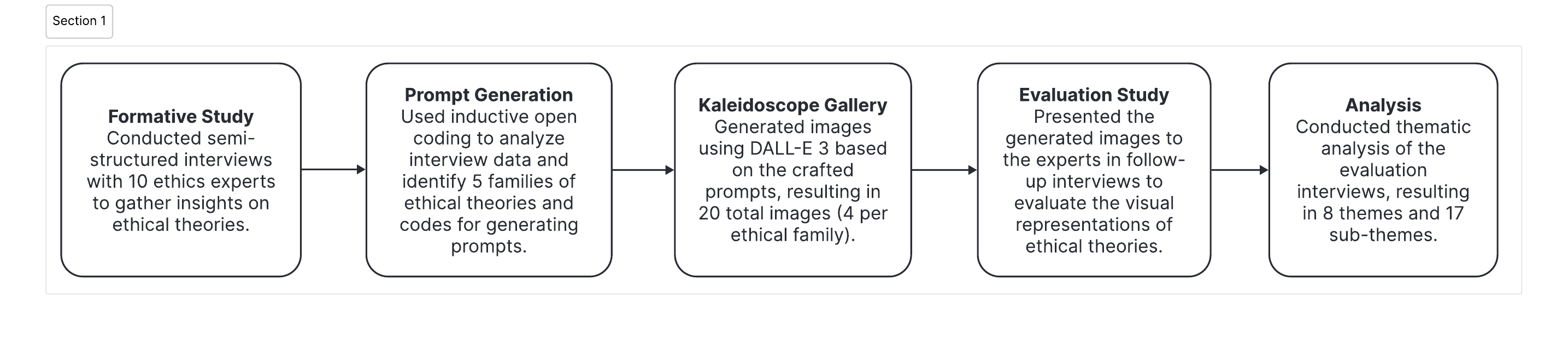}
    \caption{Summary of Study Procedure}
    \label{fig:methods}
\end{figure*}

We conducted two rounds of interviews with ten experts in ethics. In the first round, we conducted a formative study in a semi-structured interview format to elicit subject matter knowledge on ethics and introduce the metaphor of the kaleidoscope to conceptualize the ever-changing nature of both GenAI and ethics. The interview results was used to construct prompts and generate the imagery of \textit{Kaleidoscope Gallery}. In the second round, we referred back to the experts to evaluate our curation. 
All interview sessions were recorded and transcribed using Microsoft Word's transcription feature by two researchers. An iterative thematic analysis was performed on the interviews. We detail our methodology in this section, which is summarized in Figure~\ref{fig:methods} above. 

\subsection{Participants}
We used snowball sampling~\cite{snowball} to identify and recruit ten experts in ethics (5 male and 5 female). To provide a wide variety of perspectives, we deliberately recruited participants from an array of disciplines where expertise was determined by a doctoral degree or current practice in an ethics-related discipline such as philosophy and law. As shown in Table~\ref{tab:experts}, all participants encounter ethics either directly or indirectly within their discipline.

\begin{table*}[!htb]
\centering
\caption{Demographics of Interview Participants}
\begin{tabular}{cccll}
\hline
ID & Gender & Format & Academic Background & Practicing Discipline \\
\hline
P1 & F & In-person & Law & Law and Technology \\
P2 & M & In-person & Philosophy & Ethics \\
P3 & M & In-person & Philosophy & Ethics \\
P4 & M & In-person & Philosophy and Art & Arts \\
P5 & F & In-person & Philosophy and Law & Arts \\
P6 & M & Virtual & Philosophy & Ethics and Epistemology \\
P7 & F & Virtual & Computer Science & Ethical Computing \\
P8 & F & Virtual & English Literature & Ethical Technology \\
P9 & M & Virtual & Philosophy, Politics, and Economics & Ethics and Political Philosophy \\
P10 & F & In-person & Law and Sociology & Law and STS \\
\hline
\end{tabular}
\label{tab:experts}
\end{table*}

\subsection{Formative Interview Study}

\begin{figure}[!h]
    \centering
    \subfloat[\centering Mock of kaleidoscope design probe for in-person interviews. Permission was obtained from the subject for the inclusion of the image.]{\includegraphics[alt={In-person interviews of participants interacting with the kaleidoscope as the design probe.}, width=0.49\columnwidth]{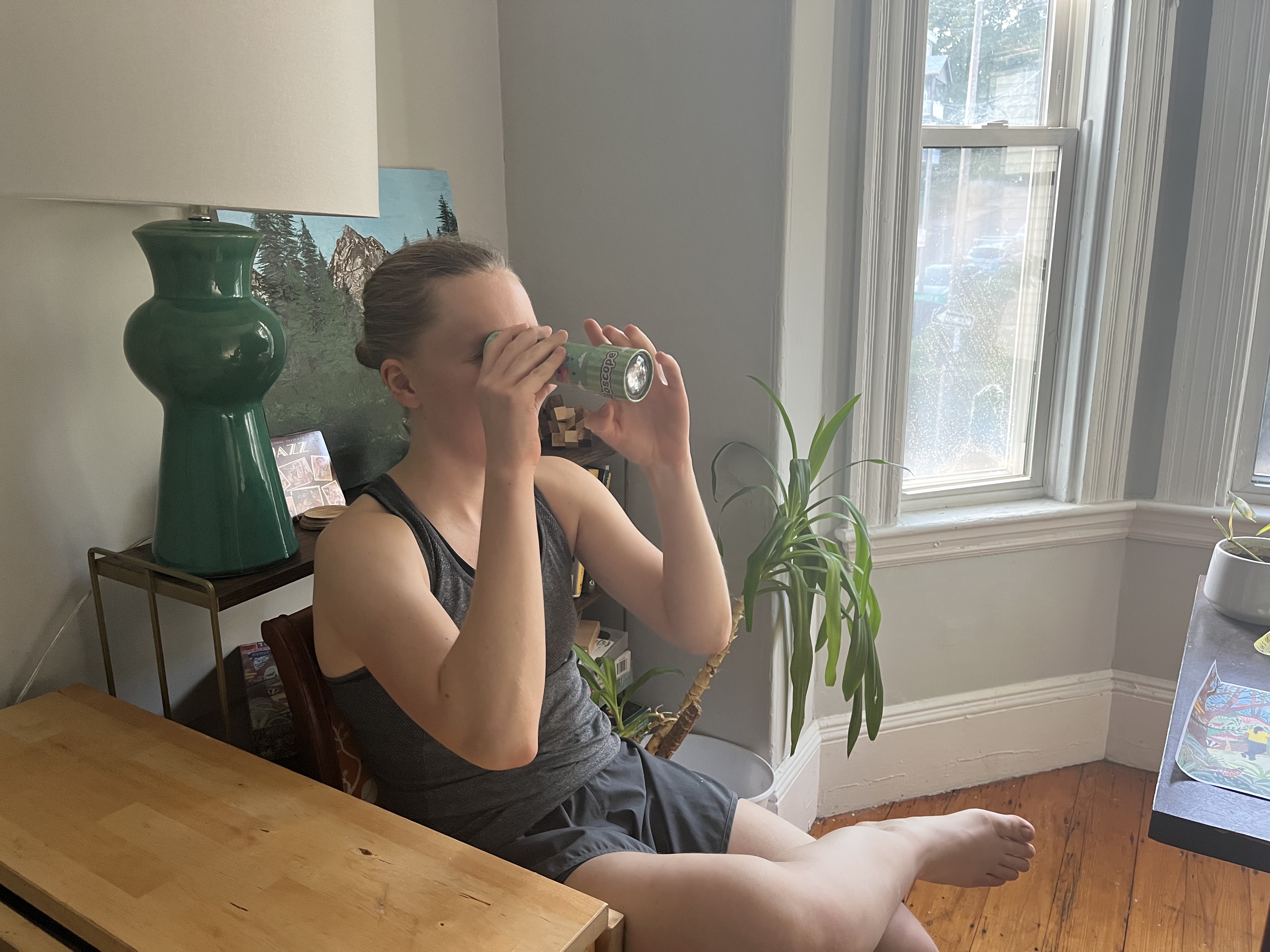} \label{fig:in-person}}
    \subfloat[\centering Virtual interviews. Participants can see shifting shapes in the rotating chamber of the kaleidoscope through the front camera.]{\includegraphics[alt={Virtual interviews of participants interacting with the kaleidoscope as the design probe.}, width=0.49\columnwidth]{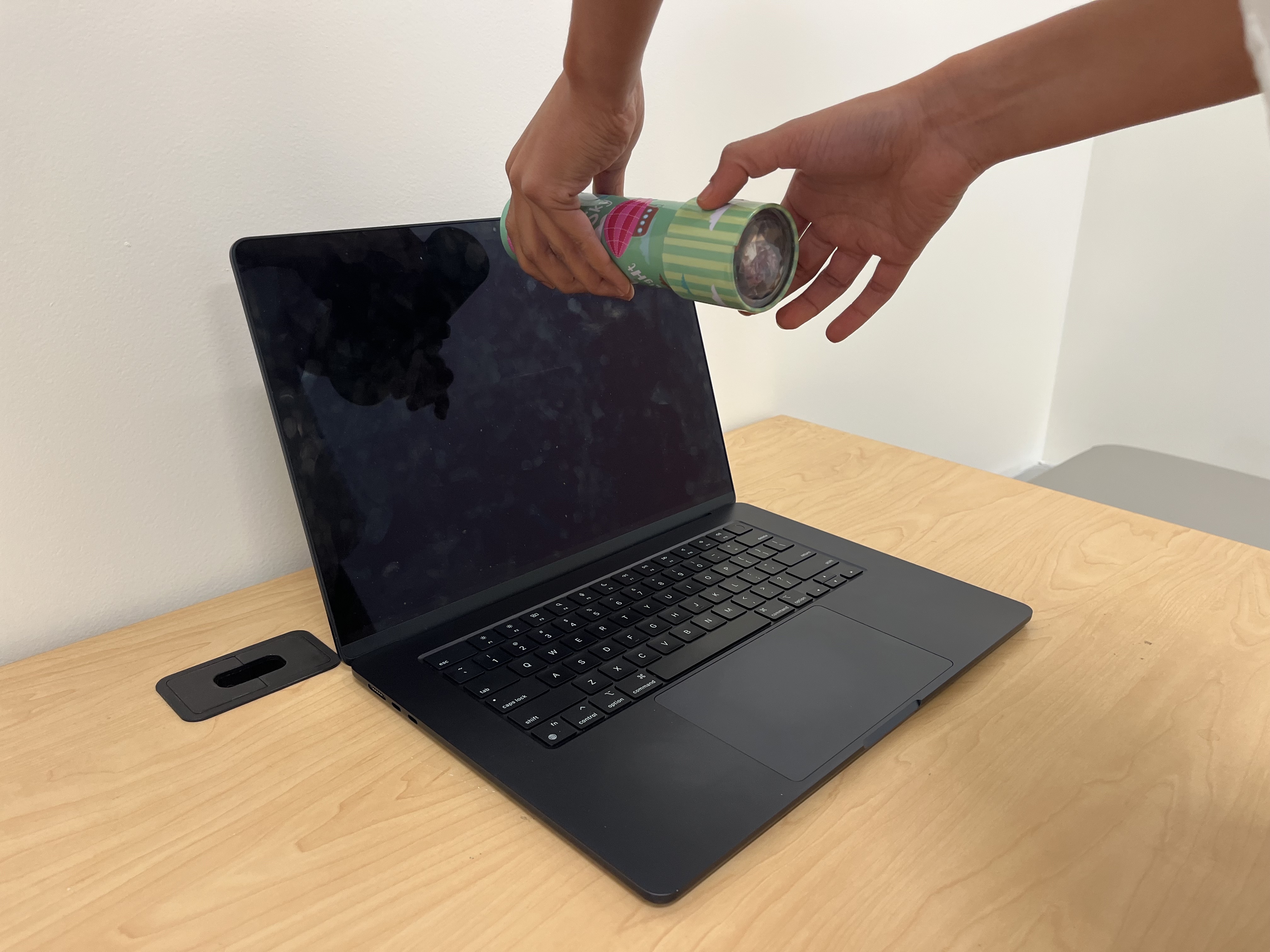} \label{fig:virtual}}
    \caption{Design probe used to onboard participants to the study and invoke their imagination.}
    \label{fig:probe}
\end{figure}

The formative study sought to gather insight and perspectives from participants on the underpinnings of ethical theories. Interviews were conducted both in person and remotely via Teams and Zoom. The interviews were semi-structured, with an average length of 30--45 minutes. Interviews began by providing the kaleidoscope as a design probe (Figure~\ref{fig:probe}). 

The probe was provided to give participants a tangible understanding of the metaphor, i.e., to look into, conceptualize, and grasp how it is primed in the study (please refer to Appendix A for the introduction of the kaleidoscope). In the interviews, participants were asked to detail their understanding of seven prominent ethical theories (utilitarian, deontological, virtue, subjectivism, care, situational, and consequences), as well as additional theories of their choosing. Participants then were asked to describe \textit{definitions} and \textit{practices} of the mentioned theories. As part of our interview process and asking about ethical theories, we purposely maintained the separation between definition and practice that is prevalent in ethics. For instance, the deontological ethical theory may tell one to follow the categorical imperative (definition), whereas its implementation (practice) would be to remain truthful at all times. The detailed interview script can be found in Appendix A. 

\subsection{Prompt Generation}

In an inductive open-coding approach ~~\cite{saldana_coding_2021}, we first performed consensus coding in two interviews to establish rigor and align coder variability~\cite{shenton_strategies_2004,richards_practical_nodate}. This was carried out by two researchers. We then separated to perform split coding for three interviews and returned for the remaining two interviews to review our overall consensus. We proceeded with a two-step process to reduce and categorize the codes using Microsoft Excel. Finally, we identified the five families of ethical theories spun from both selected and participant-introduced ethical theories. These families are Virtue, Duty-based, Consequentialism, Contractualism, and Pluralism.
We then proceeded to generate the prompts for the T2I model. Our prompts distinguish between definition and practice, in line with the consensus of ethics as theory and praxis. The descriptions of the families and the coded prompts are as follows:

\begin{enumerate}
    \item Virtue: pursuing the development of good character traits as the foundation for ethical behavior
    ~\cite{virtue_hursthouse_2023}.
    \begin{itemize}
        \item \textit{Definition prompt}: Theorizing intuitions for instinctual habits over time fostering reliable dispositions to act.
        \item \textit{Practice prompt}: Mapping of relations between oneself and role models in adopting morally good character traits.
    \end{itemize}
    
    \item Duty-based: following moral duties and principles as inherently right, regardless of the consequences~\cite{deontological_alexander_2024}.
    \begin{itemize}
        \item \textit{Definition prompt}: Codifying moral practice for inherent principles that are appropriately responsive to universalized rights and duties.
        \item \textit{Practice prompt}: Inhabiting moral abidance and code that is capable of being universalized.
    \end{itemize}
    
    \item Consequentialism: evaluating the rightness or wrongness of actions solely based on their consequences
    ~\cite{consequentialism_sinnott-armstrong_2023}.
    \begin{itemize}
        \item \textit{Definition prompt}: Framework for measuring outcome on the moral weight of conclusions advanced from a logical premises of cause and effect.
        \item \textit{Practice prompt}: Empirical analysis for the cause and effect of axioms to an applied conclusion.
    \end{itemize}
    
    \item Contractualism: basing moral principles on agreements, reason and justification
    ~\cite{contractualism_ashford_2018}.
    \begin{itemize}
        \item \textit{Definition prompt}: Codifying laws and regulations on the logical premises of coordinating society. 
        \item \textit{Practice prompt}: Systems and infrastructure to substantiate agreement and boundaries among individuals and society.
    \end{itemize}
    
    \item Pluralism: 
    accompanying multiple moral values (as equally fundamental)
    to resolve moral conflict~\cite{pluralism_mason_2023}.
    \begin{itemize}
        \item \textit{Definition prompt}: Locus for acting out of principle and reasoning out of context for the evolving nature of moral judgment. 
        \item \textit{Practice prompt}: Practical wisdom to make decisions in the pragmatic nature of geography, culture, and society.
    \end{itemize}
\end{enumerate}

\subsection{Kaleidoscope Gallery}

\begin{figure}
    \centering
    \subfloat[\centering Definition 1]{{\includegraphics[alt={Image of Virtue Ethics}, width=0.49\columnwidth]{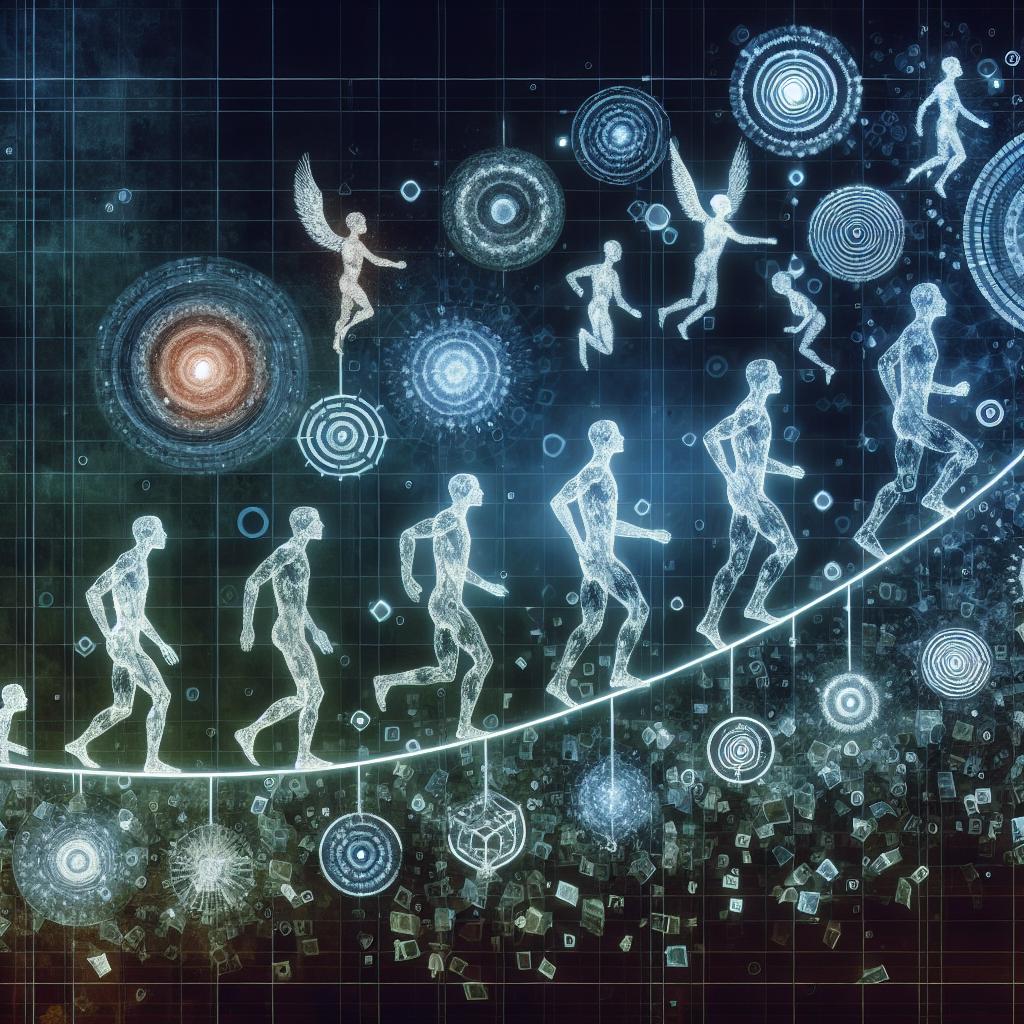}} \label{fig:vd1}}\hspace{0.01pt}
    \subfloat[\centering Definition 2]{{\includegraphics[alt={Image of Virtue Ethics}, width=0.49\columnwidth]{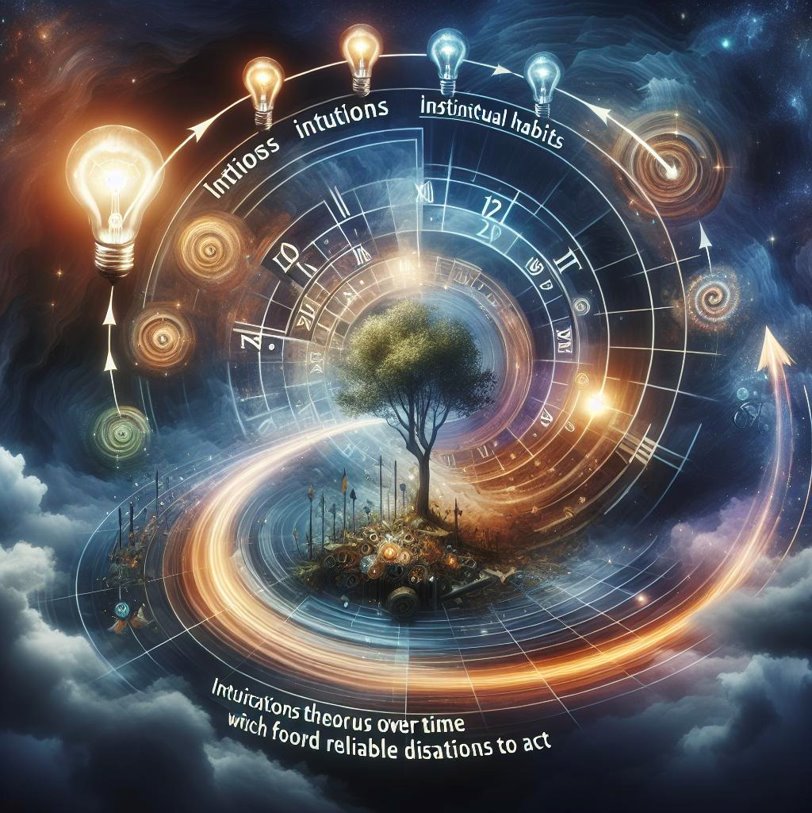}}\label{fig:vd2}}\hspace{0.01pt}
    \quad
    \subfloat[\centering Practice 1]
    {{\includegraphics[alt={Image of Virtue Ethics}, width=0.49\columnwidth]{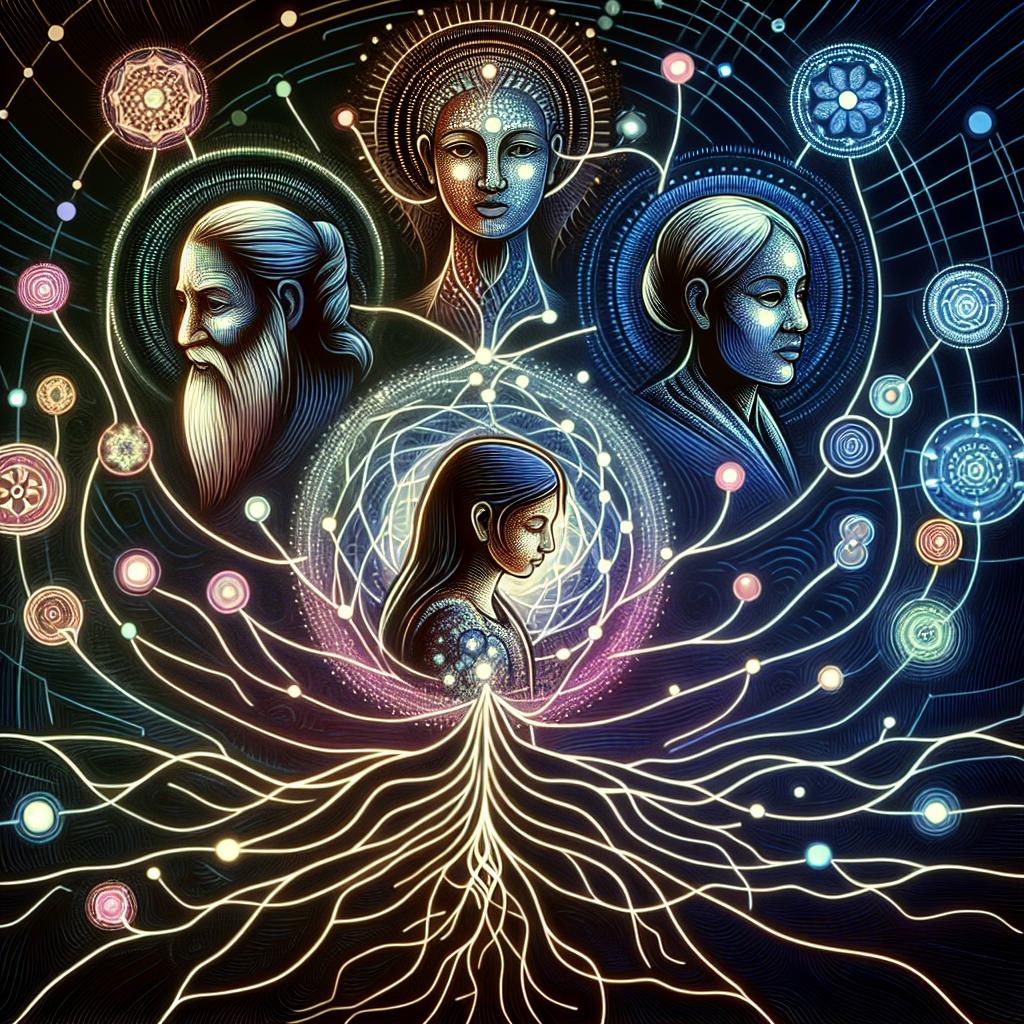}}\label{fig:vp1}}\hspace{0.01pt}
    \subfloat[\centering Practice 2]
    {{\includegraphics[alt={Image of Virtue Ethics}, width=0.49\columnwidth]{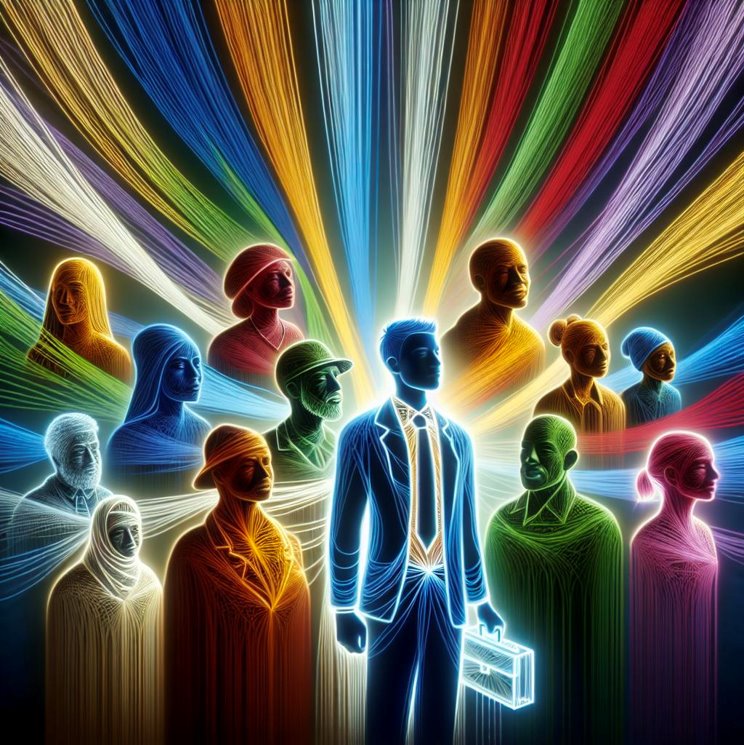}} \label{fig:vp2}}
    \caption{Virtue Ethics}
    \label{fig:virtue}
\end{figure}

\begin{figure}
    \centering
    \subfloat[\centering Definition 1]{{\includegraphics[alt={Image of Duty-based Ethics}, width=0.49\columnwidth]{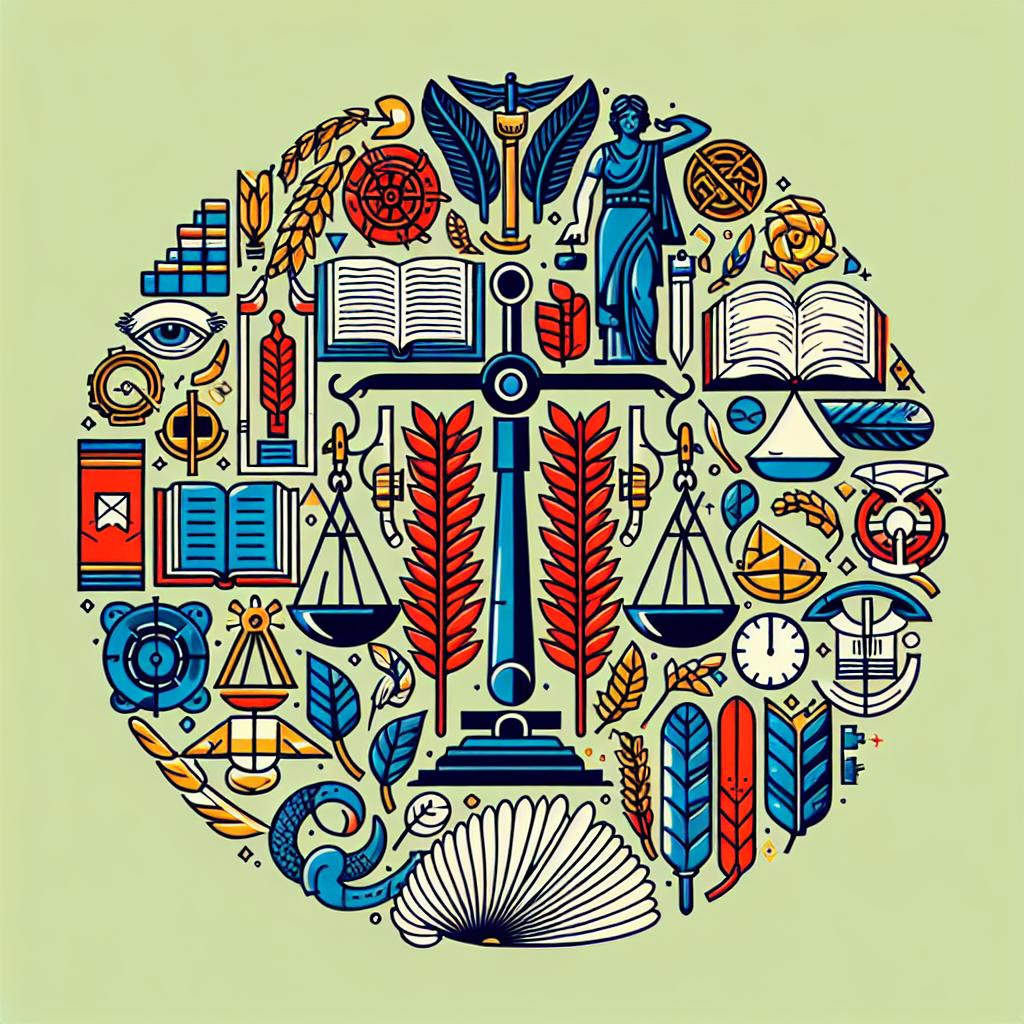}} \label{fig:dbd1}}\hspace{0.01pt}
    \subfloat[\centering Definition 2]{{\includegraphics[alt={Image of Duty-based Ethics}, width=0.49\columnwidth]{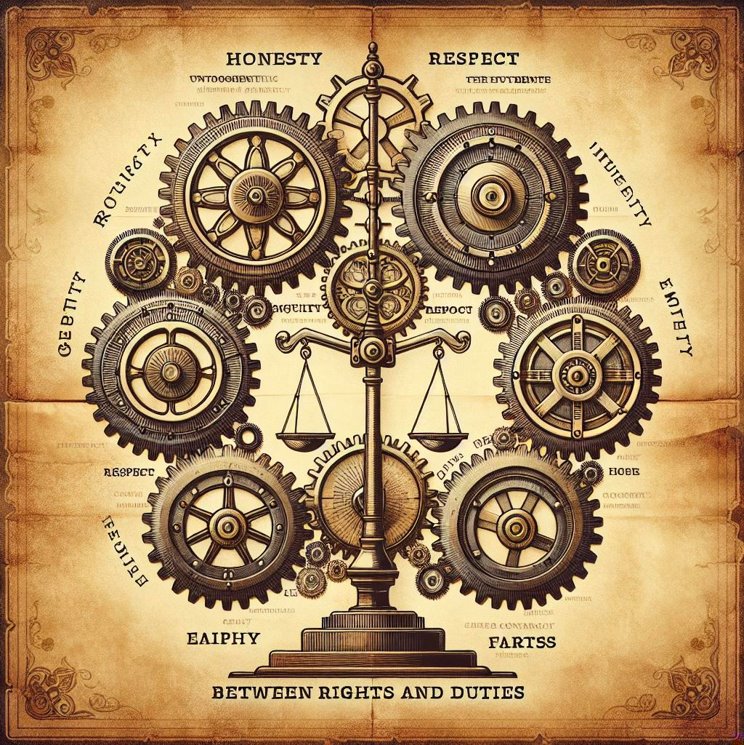}}\label{fig:dbd2}}\hspace{0.01pt}
    \quad
    \subfloat[\centering Practice 1]
    {{\includegraphics[alt={Image of Duty-based Ethics}, width=0.49\columnwidth]{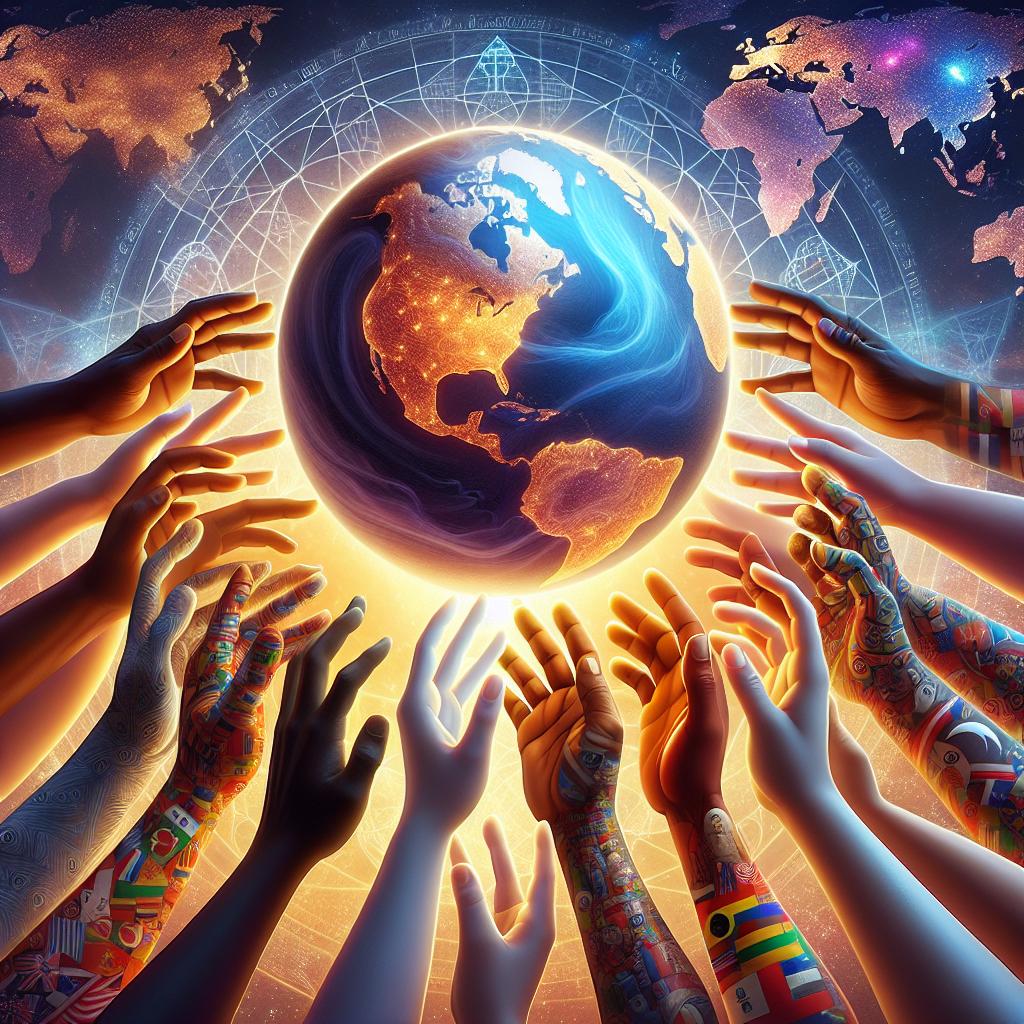}}\label{fig:dbp1}}\hspace{0.01pt}
    \subfloat[\centering Practice 2]
    {{\includegraphics[alt={Image of Duty-based Ethics}, width=0.49\columnwidth]{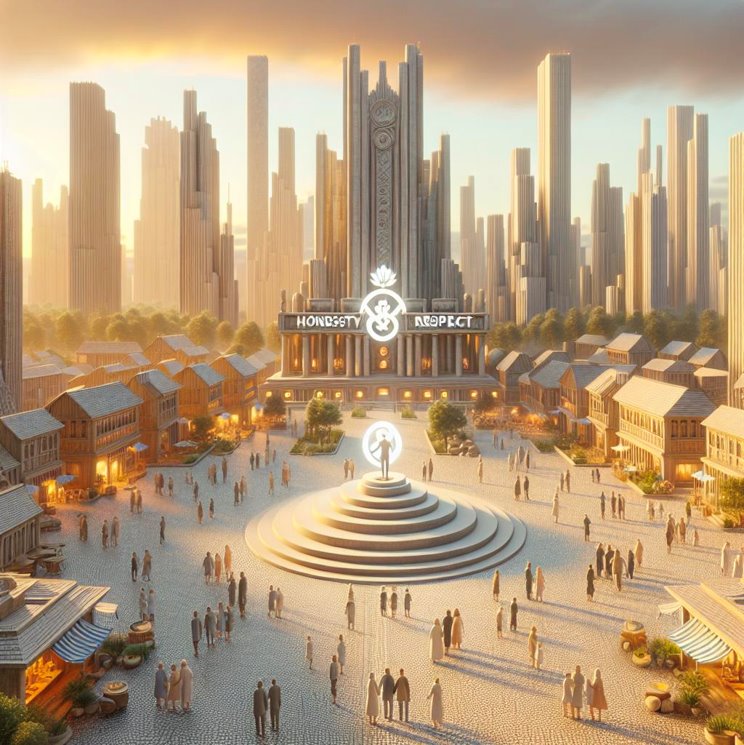}} \label{fig:dbp2}}
    \caption{Duty-based Ethics}
    \label{fig:duty-based}
\end{figure}

\begin{figure}
    \centering
    \subfloat[\centering Definition 1]{{\includegraphics[alt={Image of Consequential Ethics}, width=0.49\columnwidth]{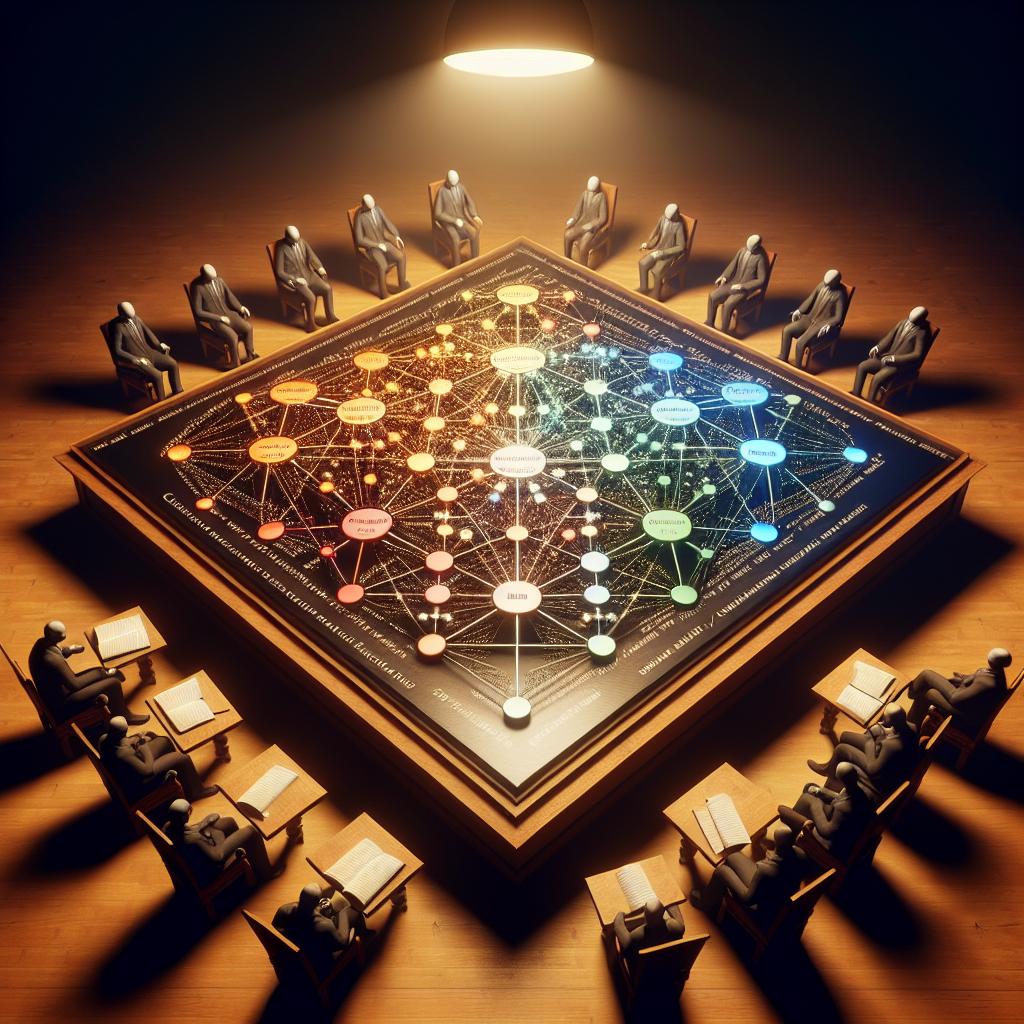}} \label{fig:consd1}}\hspace{0.01pt}
    \subfloat[\centering Definition 2]{{\includegraphics[alt={Image of Consequential Ethics}, width=0.49\columnwidth]{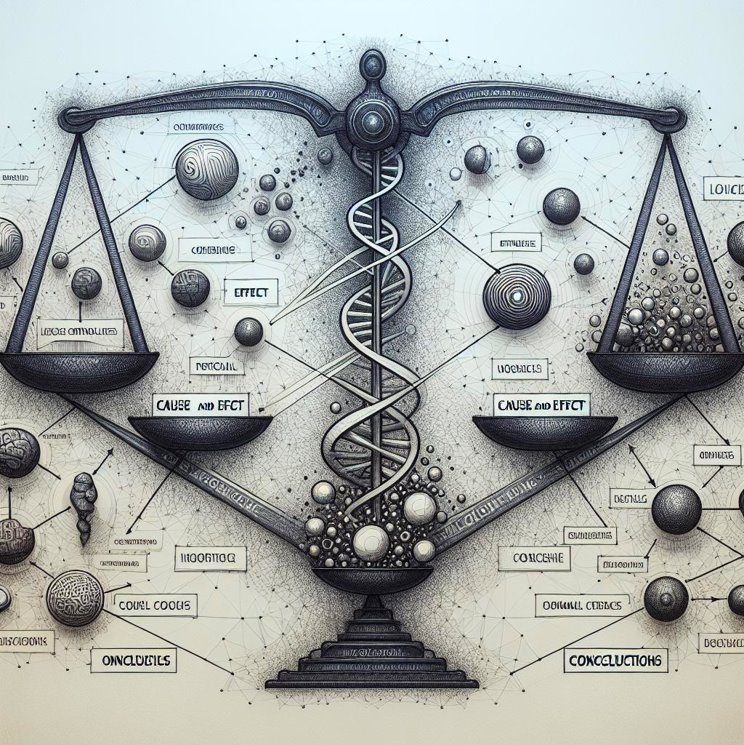}}\label{fig:consd2}}\hspace{0.01pt}
    \quad
    \subfloat[\centering Practice 1]
    {{\includegraphics[alt={Image of Consequential Ethics}, width=0.49\columnwidth]{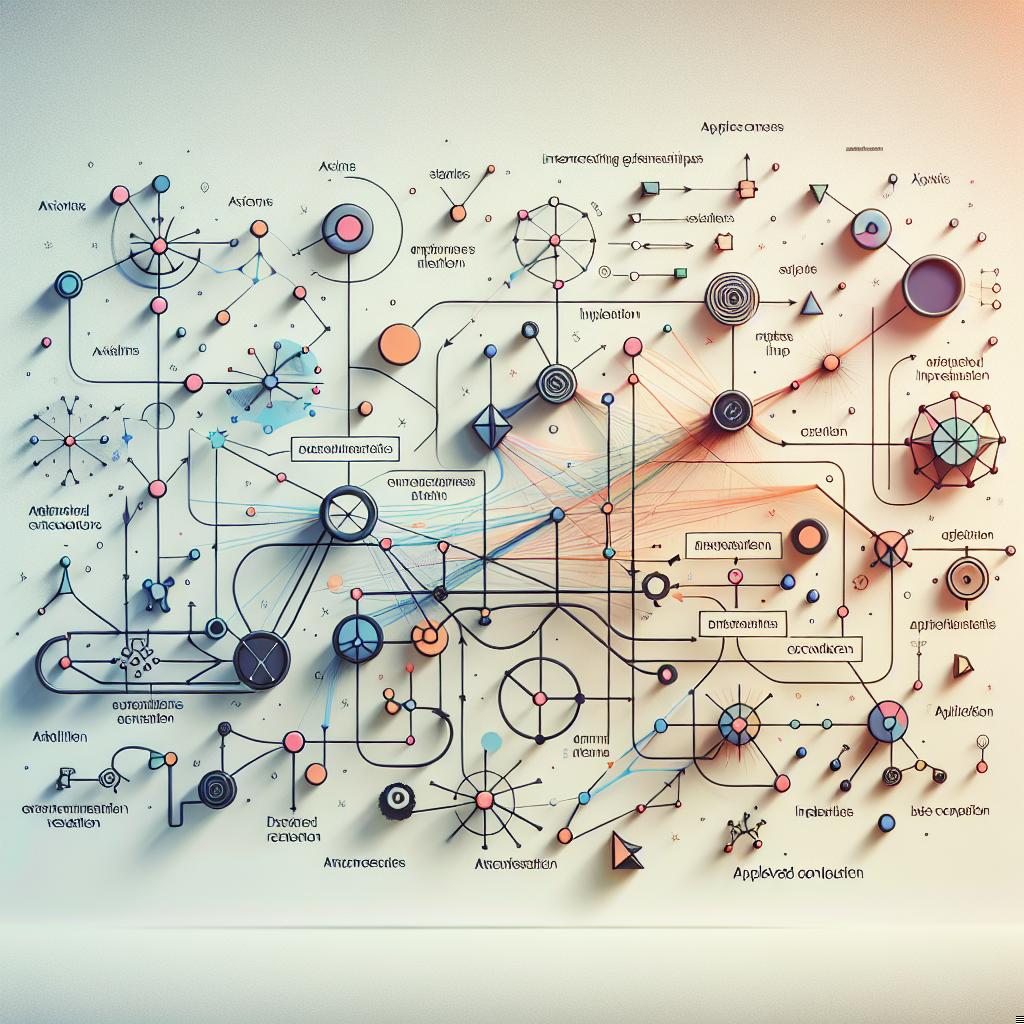}}\label{fig:consp1}}\hspace{0.01pt}
    \subfloat[\centering Practice 2]
    {{\includegraphics[alt={Image of Consequential Ethics}, width=0.49\columnwidth]{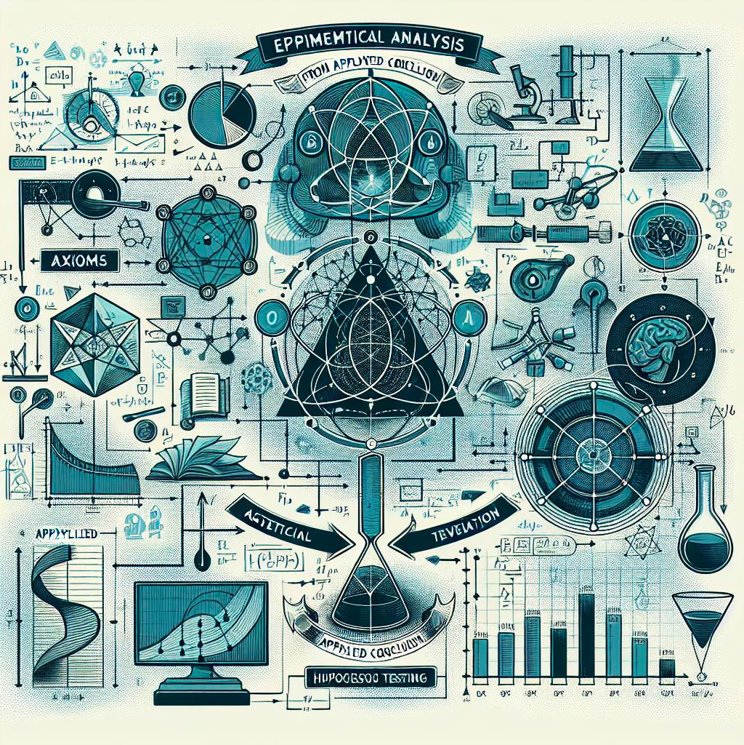}} \label{fig:consp2}}
    \caption{Consequential Ethics}
    \label{fig:consequential}
\end{figure}

\begin{figure}
    \centering
    \subfloat[\centering Definition 1]{{\includegraphics[alt={Image of Contractual Ethics}, width=0.49\columnwidth]{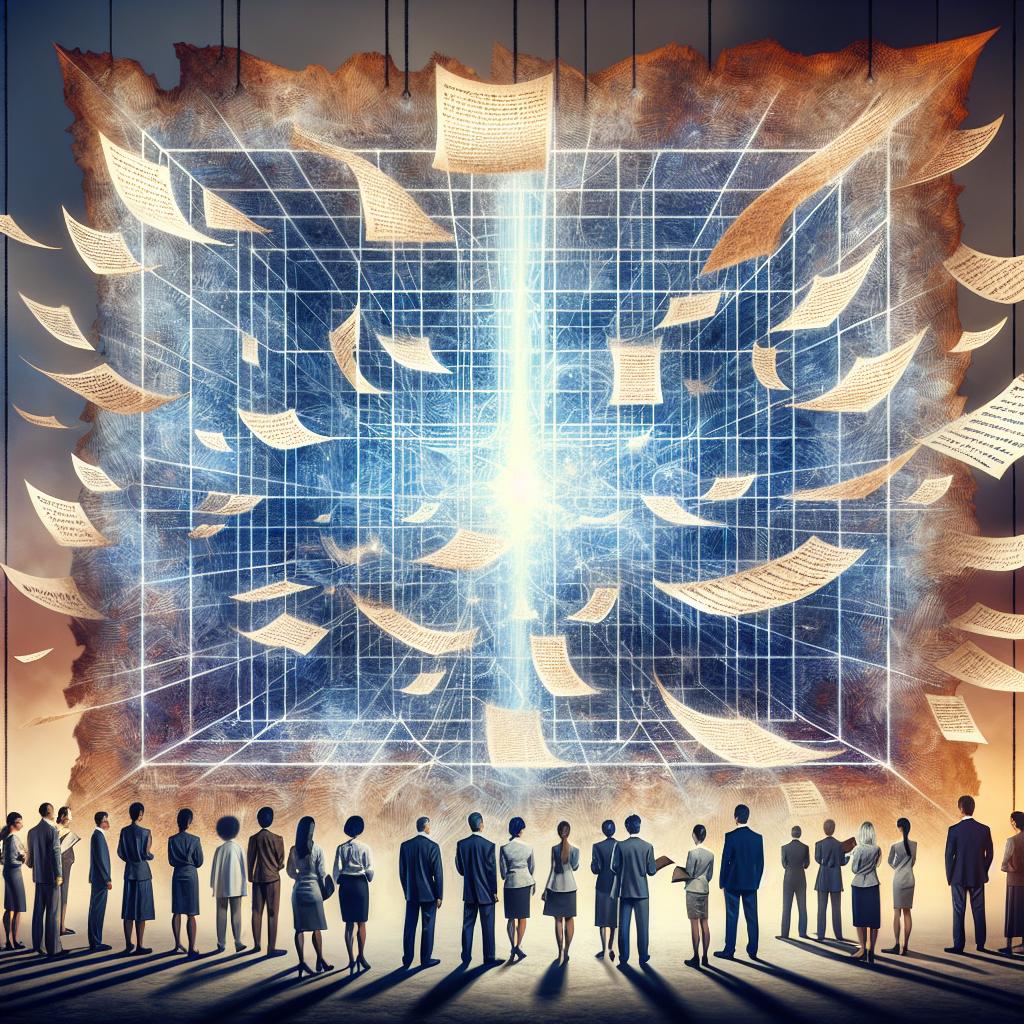}} \label{fig:contd1}}\hspace{0.01pt}
    \subfloat[\centering Definition 2]{{\includegraphics[alt={Image of Contractual Ethics}, width=0.49\columnwidth]{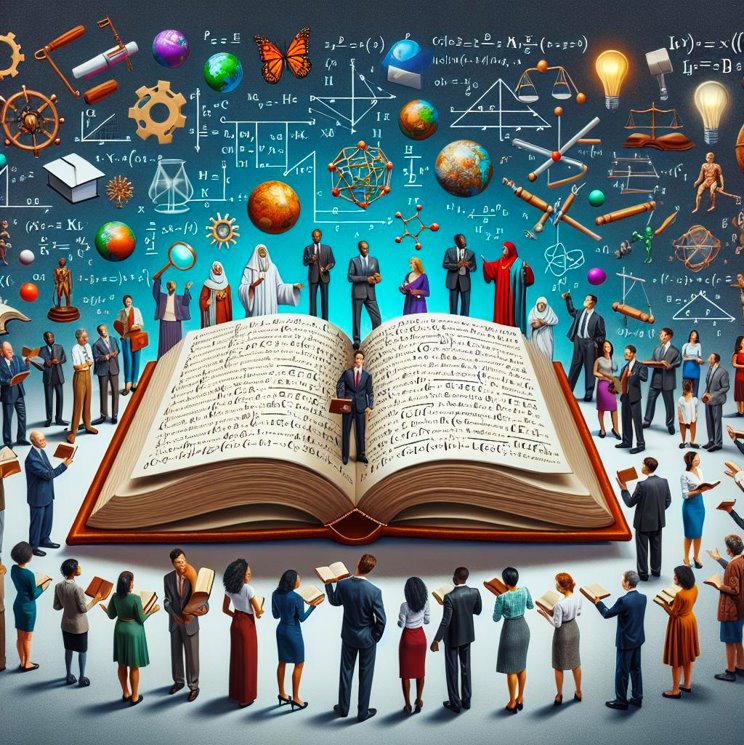}}\label{fig:contd2}}\hspace{0.01pt}
    \quad
    \subfloat[\centering Practice 1]
    {{\includegraphics[alt={Image of Contractual Ethics}, width=0.49\columnwidth]{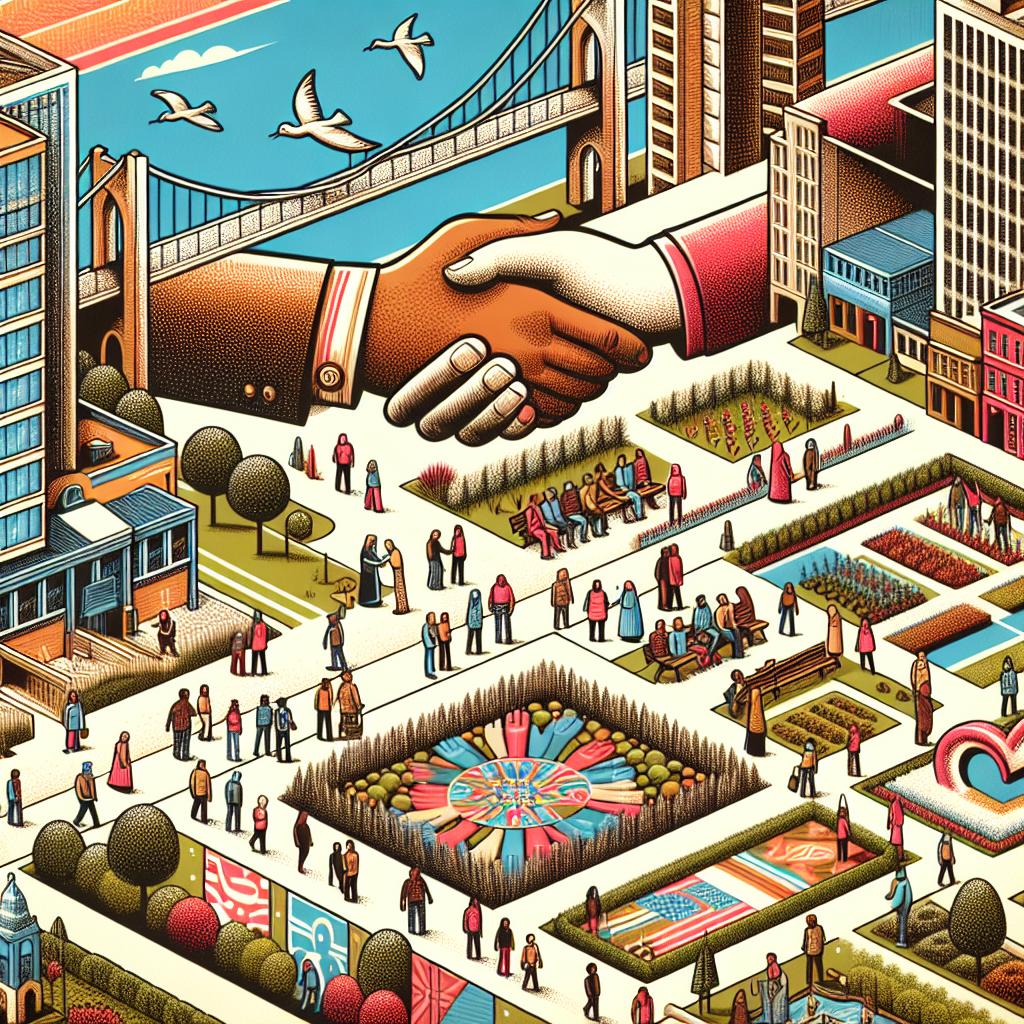}}\label{fig:contp1}}\hspace{0.01pt}
    \subfloat[\centering Practice 2]
    {{\includegraphics[alt={Image of Contractual Ethics}, width=0.49\columnwidth]{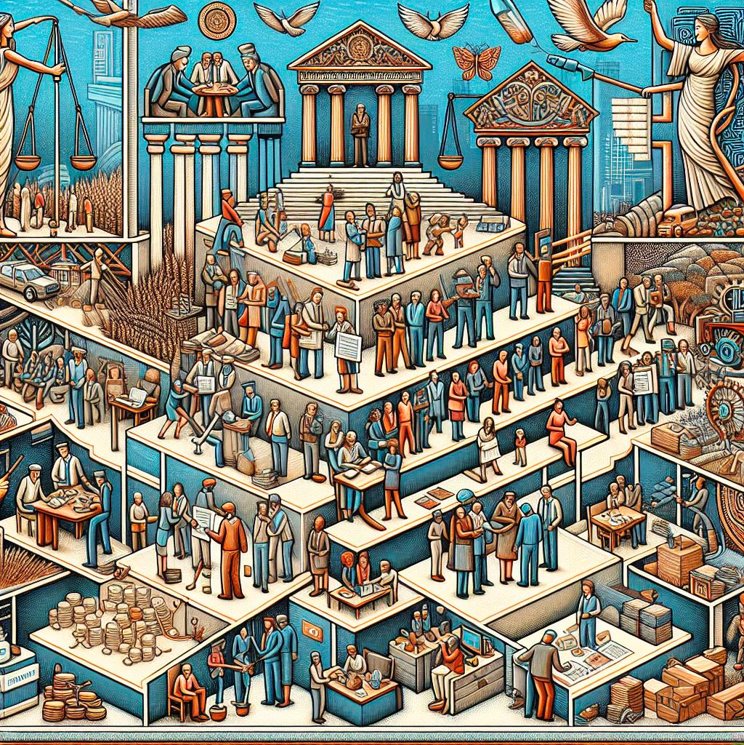}} \label{fig:contp2}}\hspace{0.01pt}
    \caption{Contractual Ethics}
    \label{fig:contractual}
\end{figure}

\begin{figure}
    \centering
    \subfloat[\centering Definition 1]{{\includegraphics[alt={Image of Pluralist Ethics}, width=0.49\columnwidth]{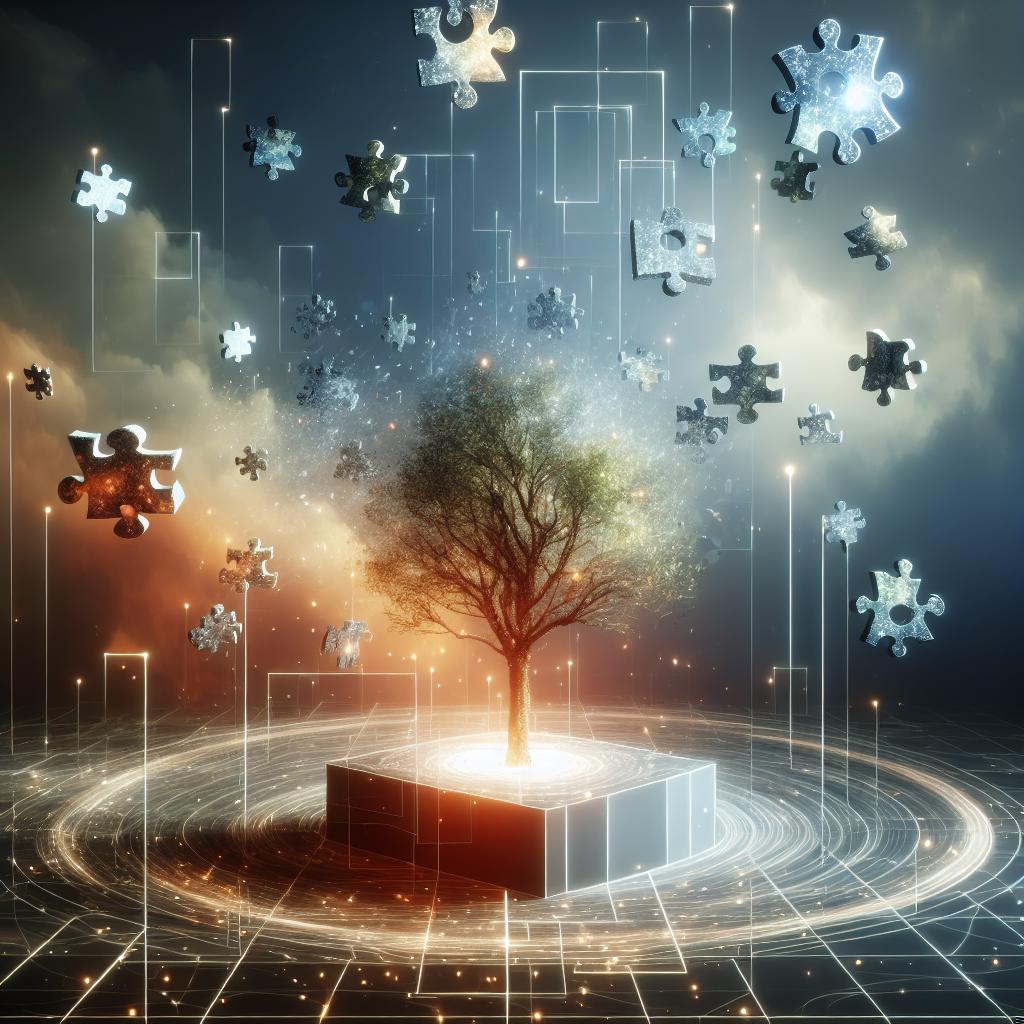}} \label{fig:pd1}}\hspace{0.01pt}
    \subfloat[\centering Definition 2]{{\includegraphics[alt={Image of Pluralist Ethics}, width=0.49\columnwidth]{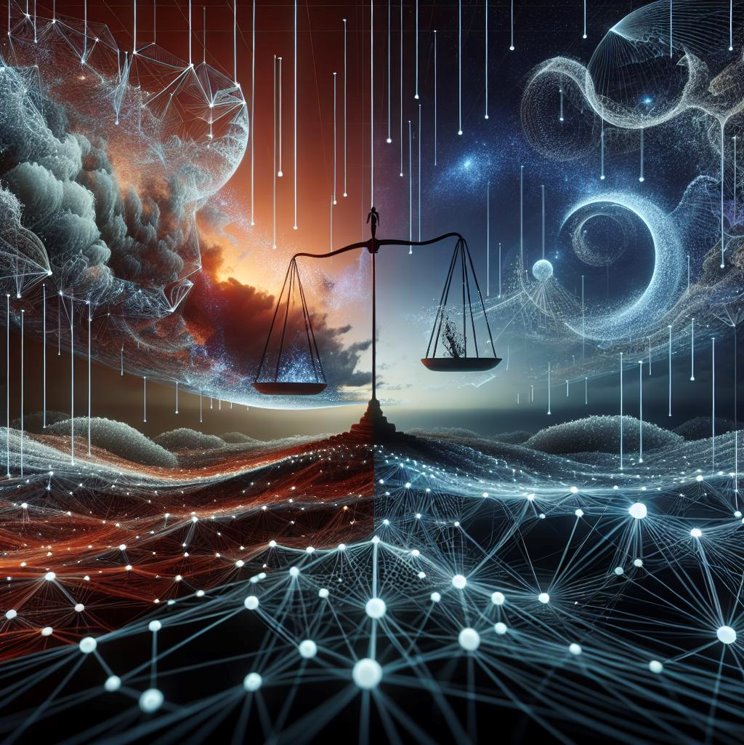}}\label{fig:pd2}}\hspace{0.01pt}
    \quad
    \subfloat[\centering Practice 1]
    {{\includegraphics[alt={Image of Pluralist Ethics}, width=0.49\columnwidth]{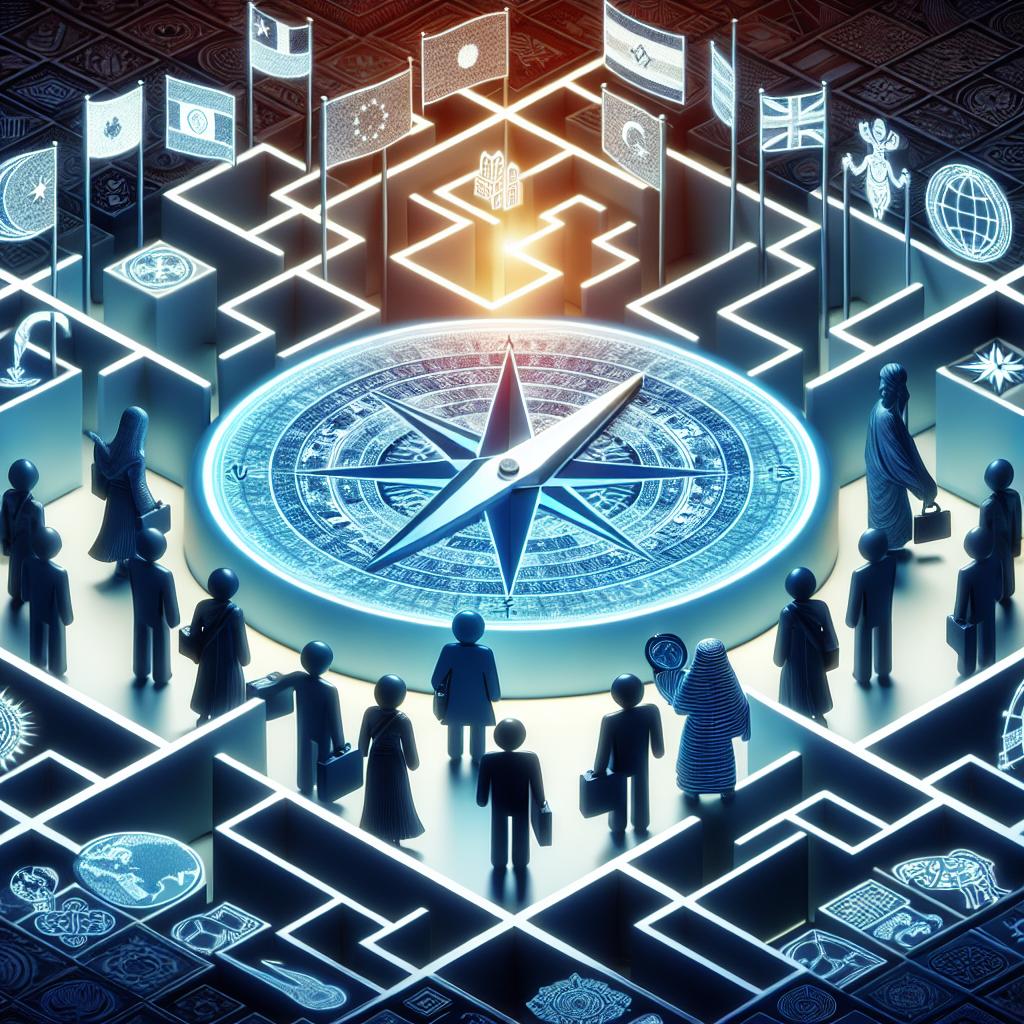}}\label{fig:pp1}}\hspace{0.01pt}
    \subfloat[\centering Practice 2]
    {{\includegraphics[alt={Image of Pluralist Ethics}, width=0.49\columnwidth]{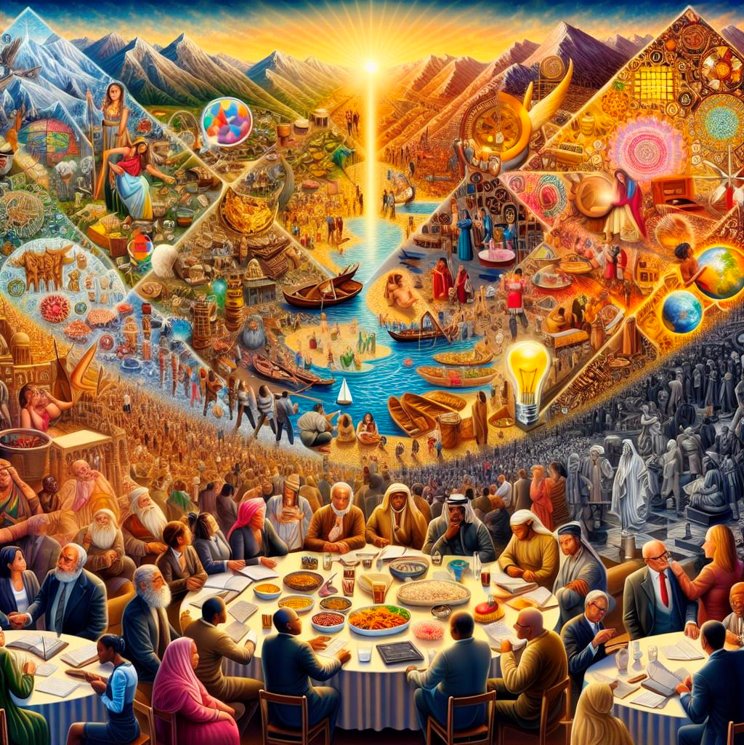}} \label{fig:pp2}}\hspace{0.01pt}
    \caption{Pluralistic Ethics}
    \label{fig:pluralistic}
\end{figure}

We used DALL-E 3 to generate our ethical imagery through a custom Python script. We chose this model as a representative of T2I models due to its unique feature on prompt following---an image's attribution to the word, ordering, or meaning of the prompt--- that is central to our investigation~\cite{betker_improving_dalle3}. 
We used this model throughout our study to reduce variability when evaluating the images although it is possible to use different T2I models. Each prompt was invoked separately by two researchers to induce the subjective nature of the model's image generation. With two images per prompt and two prompts per family, this resulted in four images per family and a total of twenty for the curation of \textit{Kaleidoscope Gallery}. We used four images per ethical family in order to give experts an insight into the kaleidoscope metaphor without crowding their visuals and ability to pick up details of the generated images. We also chose two images per prompt to track discrepancies in image generation for images generated from the same prompt. We display our image generations in Figures ~~\ref{fig:virtue}--\ref{fig:pluralistic}.

\subsection{Evaluation Study}

\begin{figure}[!h]
  \includegraphics[alt={Exhibit of Kaleidoscope Gallery shown to participants. This slide represents Virtue Ethics}, width=\columnwidth]{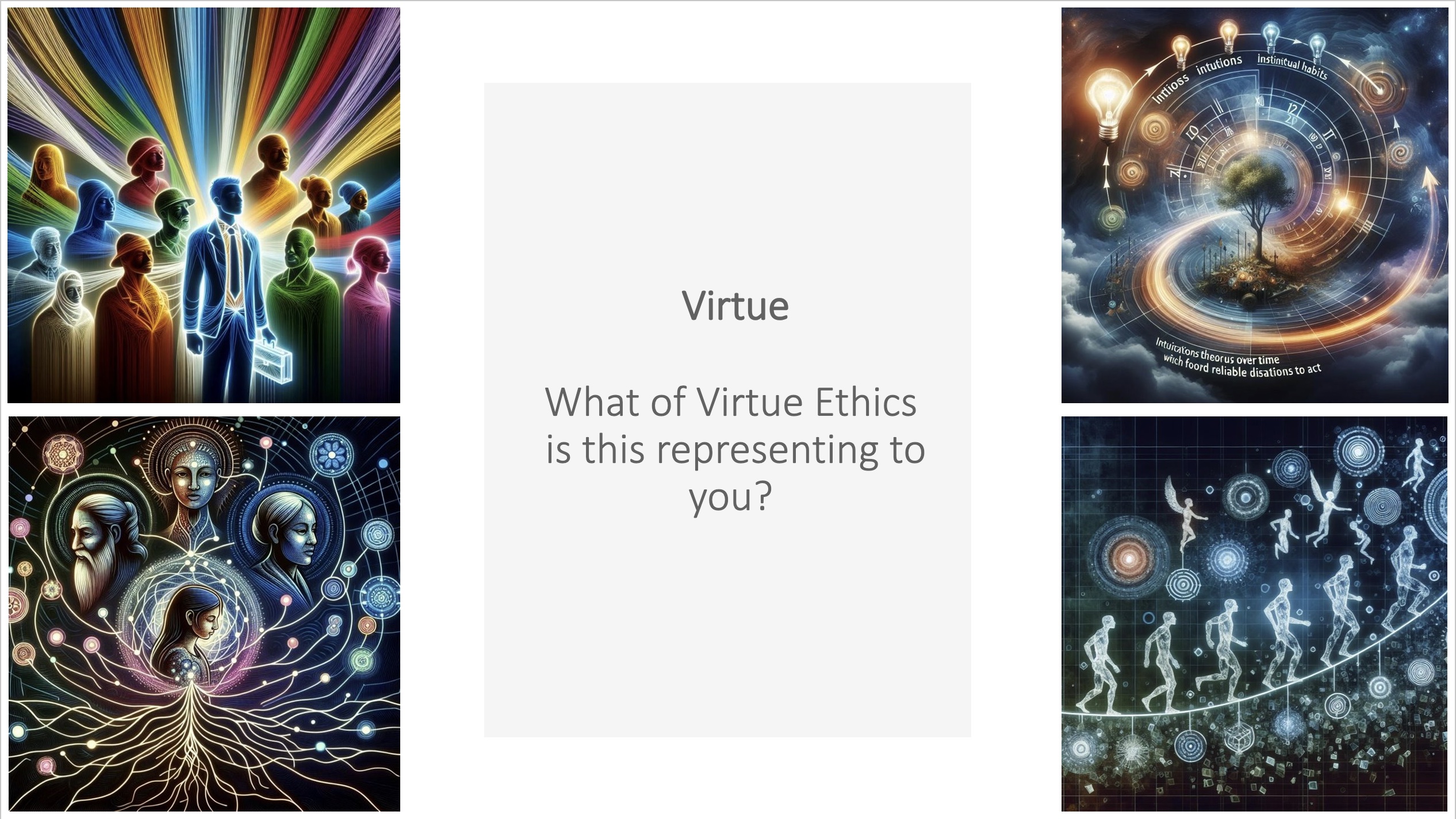}
  \caption{Exhibit of \textit{Kaleidoscope Gallery} for Virtue Ethics}
  \label{fig:exhibit}
\end{figure}

The generated images were grouped according to their corresponding ethical family and presented to the experts in a 30-minute follow-up interview showcasing \textit{Kaleidoscope Gallery}. Figure~\ref{fig:exhibit} presents a slide that was shown to participants for Virtue Ethics. All interviews were conducted remotely via Zoom or Teams and were video-recorded for transcription. 

In the exhibit, we did not show the experts the prompts used to generate the images because we did not want to bias their interpretations of the images. Hence, the images for definition and practice were presented together. Similarly, each slide was exhibited with the same question tailored to the related ethical family. Upon conducting the evaluation study, we used a three-step iterative coding process consistent with the approach used in our formative study to analyze our video transcriptions. The final thematic analysis of our code book found three categories, eight themes, and seventeen sub-themes that reflect experts understanding of ethics as they asses the generated images (RQ3). We acknowledge that there could have been biases outside the scope of the direct presentation of images (curation of Kaleidoscope Gallery) in our work. These include the biases that are likely present in DALL-E 3, the biases that the participants have towards GenAI that might impact their evaluation and the biases perpetuated in the use of GenAI and images created. We discuss these potential biases in the limitations of our work.

\section{Results}

We summarize our results in Table~\ref{tab:results}. Our themes highlight how morality (S4.1, S4.2), society (S4.3, S4.4, S4.5), and learned associations (S4.6, S4.7, S4.8) are central to ethical theories. We detail each theme in this section.

\begin{table*}[h]
    \centering
      \caption{The three categories, eight themes, and corresponding sub-themes developed through our analysis.}
    \begin{tabular}{cccccc}   
         \hline
         Category & S & Themes & Sub-themes & Count\\
         \hline
         \multirow{4}{*}{Morality} & 4.1 & Navigating Moral Compass & Internal Dialogue & 7\\
         & & & Growth Path & 7\\
         \cline{2-5}
         & 4.2 & Representation of Morality & Cosmic Imagery & 5\\
         & & & Moral Recognition & 5\\
         \hline
         \multirow{7}{*}{Society} & 4.3 & Bias of Representation & Gender Bias & 4\\
         & & & Geographic Bias & 4\\
         \cline{2-5}
         & 4.4 & Difference and Diversity & Multicultural and Multiplicity & 7\\
         & & & East–West dichotomy & 6\\
         \cline{2-5}
         & 4.5 & Social Construct & Structure and Governance & 9\\
         & & & Hierarchies of Governance & 7\\
         & & & Cooperation and Collaboration & 9\\
         \hline
         \multirow{6}{*}{Learned Associations} & 4.6 & Decision-making Procedures & Frameworks & 9\\
         & & & Mechanisms & 9\\
         \cline{2-5}
         & 4.7 & Signs and Symbolism & Universality & 6\\ 
         & & & Navigation & 5\\
         \cline{2-5}
         & 4.8 & Image Composition & Geometry & 7\\
         & & & Color & 6\\
         \hline
    \end{tabular}
    \label{tab:results}
\end{table*}

\subsection{Navigating Moral Compass}

This theme emphasizes the importance of internal processes and personal development in navigating one's moral compass. \textit{Internal dialogue} serves as a tool for self-examination and decision-making, while \textit{growth path} represents the ongoing journey of moral improvement and self-realization.

\subsubsection{Internal Dialogue} Concepts such as individuality, telos (ultimate purpose), suffering, principles, and the evolution of the self
gave way to an internal dialogue for moral growth and development. The images that inspired these reflections include the tree in Figure~\ref{fig:vd2} (virtue ethics) for a person ``cultivating values in pruning to a particular direction'' (P2) and Figure~\ref{fig:vd1} (virtue ethics) for persons finding ``their better angels'' (P6). Participants gave examples addressing \textit{questions to the self} framing prompts such as ``What am I doing? What are the implications going to be and how do I weigh them up against one another?'' (P3), ``What kind of person do they want to be?'' (P4), and ``How do you identify which values to follow when they are equally applicable and might be conflicting?'' (P8). A model of the mind was also analyzed as a visual representation of such internalization. P6 noted Figure~\ref{fig:dbd2} (duty-based ethics) to depict ``the mind through its interaction of different parts" in navigating a moral landscape. 

\subsubsection{Growth Path} Equally prevalent was the visual path of internal growth and development.
This observation was prevalent in virtue ethics and presented in the images with the idea of moral exemplars (Figure~\ref{fig:vp1}), the aspirations for striving for betterment (Figure~\ref{fig:vd1}), and the life progress in work and family (Figure~\ref{fig:vp2}). The growth of a single human life was also highlighted as a prominent attribute among most of our participants, where these actions were internal to the pursuit of attainment. However, P8 noted this path was not linear as ``actions in service of others with the expectation to do better yet with the reality of suffering and depriving yourself'' was part of the moral compass that individuals were not aware of.


\subsection{Representations of Morality}

This theme contrasts the multifaceted nature of morality, encompassing both cosmic, universal aspects and human-centric, social dimensions. As evaluated by our participants, abstract and direct interpretation with \textit{cosmic imagery} to the former and \textit{moral recognition} to the latter were prevalent representations of morality.

\subsubsection{Cosmic Imagery} In abstract representation, cosmic systems exhibited a distinctive aesthetic that was associated with morality. The systems symbolized what is ``less human-centric'' (P4) and what is ``greater than their intention'' (P5). P5 highlighted this phenomenon of morality to be natural, ``fluid'' as opposed to the ``duality'' of an absolute. Elements that signaled this synthesis include the blending of the sun and the moon, and the star system mapping the tree in Figure~\ref{fig:pd2} (pluralistic ethics). Cosmic imagery was equally mentioned alongside spiritual undertones. The angel, clouds, sky, and climbing man on a ``stairway to heaven'' (P5) in Figure~\ref{fig:vd1} (virtue ethics) and concentric energy around the tree in Figure~\ref{fig:pd1} (pluralistic ethics) took precedent to this observation. Drawing consensus on a visual theme, the central tree in Figure~\ref{fig:vd2} (virtue ethics) and~\ref{fig:pd1} (pluralistic ethics) was also equated to the religious embodiment of the ``Tree of Life'' among participants (P3, P4, P10). 

\subsubsection{Moral Recognition} Morality attained direct representation on the question of ``is it possible we still want to interact with each other on the basis of recognizing each other's humanity?'' (P2). To such questions, drawing all the theories together, answers consolidated morality to exist among virtue, duty, the legal system, agreements, and a diversity in perspective with humans as the focus of morality. 
In this sense, P8 mentioned that ``any serious moral framework is inexhaustible'' and that what is governed by the human mind is subject to ideal representation. Such imagery included  
the lack of nature in Figure ~\ref{fig:dbp2} (duty-based ethics)---nothing to be tamed in governance---as well as the stipulation of contracts in contractual ethics. P9 provided Figure~\ref{fig:contp1} (contractual ethics) as an example of being freely and equally placed in an ideal and specialized way as ``what people would agree to under certain idealized conditions''. 

\subsection{Bias of Representation}

This theme surfaces biases in representation within ethical imagery with a focus on \textit{gender} and \textit{geographic} bias.  

\subsubsection{Gender Bias} Gendered representation was a prominent issue highlighted among some participants. The male-centered depictions in virtue ethics (Figures~\ref{fig:vd1} and~\ref{fig:vp2}) were scrutinized for cultivating the idea of a moral exemplar in the locus of masculinity. P2 highlighted this observation to be related to classical virtue traditions (Buddhism, Christianity, Confucianism), where the patriarchal history and moral exemplars of the traditions lead the virtues to be associated with masculine traits such as physical courage as opposed to emotional care. Gender bias was also mentioned for the lack and imbalance of female representation. Participants observed Figure~\ref{fig:consd1} (consequential ethics) to highlight a council of male-centered governance and the idea of ``men in power.'' P5 mentioned women have ``skirts above their knees'' in Figures~\ref{fig:contd1} and~\ref{fig:contd2} (contractual ethics) in a manner that is offensive and NSFW (not safe for work) if used in the context of education. 

\subsubsection{Geographic Bias} Image generations were criticized for their geographic placement in the Western world. In Figure~\ref{fig:dbp1} (duty-based ethics), participants highlighted the Americas to be placed in the foreground with the ``rest of the world in the background'' (P2). Participants added to this analysis in the ``US-centered'' notion toward a sense of ``global duty bias'' (P5) that is not partial to the entirety of the world. In the societal depiction of pluralistic ethics (Figure~\ref{fig:pp2}), participants scrutinized the image to reinforce preconceived norms and stereotypes about global cultures and society, such as idealized livelihoods and conceptions of the roles members play in their societies (P5, P6, P9).


\subsection{Difference and Diversity}
This theme showcases how differences in understanding the world were eluded by \textit{multicultural and multiplicity} in worldviews and contrasting cultural frameworks of \textit{east–west dichotomy}.

\subsubsection{Multicultural and Multiplicity}

The observation that society has different worldviews was elicited by the multicultural representation, where people were depicted from different countries, situations, and landscapes.
These were noticed in the inter-generational figures present in Figure~\ref{fig:vp2} (virtue ethics), the makeup of society in Figure~\ref{fig:contp2} (contractual ethics), and the hands that reach the globe in Figure~\ref{fig:dbp1} (duty-based ethics). Likewise, diversity among multicultural representations and multiplicity were noticed among the conglomerations of persons. In pluralism, this resonated in Figure~\ref{fig:pp2} as ``different kinds of life, colors, people, places living together'' (P3) adjoined with the flags around the world in Figure~\ref{fig:pp1}. More so, the diverse representation was understood twofold. First, it entailed ways of prioritizing values, and second, the prioritization of values themselves---the latter was associated with the multiplicity of values and why different factors exist in the first place (P2, P6, P9). This observation was pivotal in understanding the pluralist ethical theory.

\subsubsection{East–West dichotomy} Participants noted a polarity between depictions that signaled Eastern and Western traditions. This was prominent in virtue ethics among Figures~\ref{fig:vp1} and~\ref{fig:vp2}, with the former signaling the East in a spiritual guide and the latter signaling the West in the evolution of the character (P10). This duality was also highlighted with the naturalistic depiction of the \textit{Tree of Life} in Figure~\ref{fig:vd1} (virtue ethics) as an addition to Eastern depictions. Some participants also recognized a Western framing and ``associations we already have in our culture'' (P5). This included the resemblance of the scale to represent justice and the composition of society in Figures~\ref{fig:consd2} (consequential ethics) and ~\ref{fig:contp2} (contractual ethics) among a specific locus. Participants signaled how this gave way to understanding how the model was trained (P6, P10). P2 added that they were surprised not to see depictions of animals as ``a lot of the moral life for humans also involves interaction with the world that's not just human'' suggesting that models trained on holistic datasets would depict more inclusive worldviews.



\subsection{Social Construct}

This theme illustrates how social constructs shape governance and societal organization through collective agreement and shared meanings. The sub-themes of \textit{structure and governance, hierarchies of governance} and \textit{cooperation and collaboration} reveal human interaction, symbolic meaning-making, and negotiated power dynamics that vary across ethical families.

\subsubsection{Structure and Governance} 

Social structure in the form of cities and urban organizations, such as large-scale civic institutions, provided a lens through which participants observed a world ``responding to the material foundation of idea production'' (P5). Participants noted the symbols of math, the compass, and books in Figure~\ref{fig:contd2} (contractual ethics) to be ``products of social cooperation'' (P2, P6). Order was crucial for ``higher-order knowledge'' which flowed from ``establishing peace in society'' (P10), paralleled by the idea of systems structuring governance (P3, P5, P6). A modern fantasy of utopia where freedom is taken for safety and order of governance was also described as an ordered universe. Participants detailed that the need for group governance entailed a response to the expansion of cities, a form of engagement that would address harmony among interactions between two strangers. Ethical theories as such arose for situations where ``people mostly don't go over each other'' (P2), such as a well-ordered society governed upon duty (Figure~\ref{fig:dbp2}).

\subsubsection{Hierarchies of Governance}

Triptych imagery (a three-sectioned work of art) was depicted in social structures as a pyramid of society. In Figure~\ref{fig:contp2} (contractual ethics), the working class is placed at the bottom, the middle class in the center, and the higher-level decision-makers in the display of classical antiquity (P5). In Figure~\ref{fig:dbd1} (duty-based ethics), the farmer, the worker, and the intellectual per the imagery of grain, tools, and books presents attributes that convey these levels. In Figure~\ref{fig:consd1} (consequential ethics), a two-level hierarchy was also noted with a ``technocratic'' elite  from above that makes decisions for recipients (P5, P9, P10). 
On the other hand, a juxtaposition between humans and nature served as a hierarchy between balance and order. Participants observed the presence of nature to be aligned with the ethical family -- the presentation of tamed botany in duty-based ethics (Figure~\ref{fig:dbp2}) and consequential ethics (Figure~\ref{fig:contp1}), and free-flowing nature in virtue (Figure~\ref{fig:vd2}) and pluralist ethics (Figure~\ref{fig:pd1}). P10 highlighted 
the lack of animals and children, but rather adults, to depict the awareness of choice and duty, as well as a higher level of organization that requires ``voluntarily adhering to a social order''.

\subsubsection{Cooperation and Collaboration}

Togetherness was a theme witnessed through the elements of agreement and community based on ``mutual understanding to enable cooperative living'' (P3). 
Agreement enlisted various depictions, such as agreement upon duty, conditions, causal relations, permission, and discussion. The most evident representation was the shaking of hands in Figure~\ref{fig:contp1} for contractual ethics (P4, P5, P6, P7, P8). Agreements also expanded beyond what ought to be written, such as those in a book (the book in Figure~\ref{fig:contd2} was understood to be a constitution), but rather a pledge ``acted irrespective of returns'' (P6, P8, P9). 
On the other hand, community was observed in the conveyance of a council in Figure~\ref{fig:consd1} (consequential ethics), 
family and ancestry in Figure~\ref{fig:vp1} (virtue ethics),
co-existence and harmony in Figure~\ref{fig:contp1} (contractual ethics)
, and diverse communion in Figure~\ref{fig:pp2} (pluralistic ethics)
. In pluralistic ethics, Figure~\ref{fig:pp2} was furthered to represent ``not just tolerance but acceptance'' (P6) 
whereas, Figure~\ref{fig:dbp2} in duty-based ethics showcased citizens going about their means, in that ``this is a place for them to live'' (P3). This insight informed forms of social constructs that were variant among ethical families (P2, P5, P6, P10).

\subsection{Decision-making Procedures}

This theme highlights the diversity in ethical decision-making procedures. \textit{Frameworks} emphasizes four modes of decision-making such as balance and prioritization, while \textit{mechanisms} integrate symbolic tools such as scales and networks to represent justice and moral landscapes.

\subsubsection{Frameworks} 

Decision-making procedures encountered four modes of measurement, namely balance, prioritization, scientific method, and weighing. The gear or mechanical cogs in Figure~\ref{fig:dbd2} (duty-based ethics) served as a primary indicator of balance. 
Prioritization was observed through the actions showcasing obedience to societal roles. This was mentioned in Figure~\ref{fig:dbd2} (duty-based ethics) with ``prioritizing the process where everyone is doing their job'' (P7). 
The scientific approach provided systems of ``translating or reducing situations, values, reserves into things can be computed to and calculated '' (P2) and was primarily witnessed in Figure~\ref{fig:consp2} (consequential ethics) as a scientific attitude toward morality (P1, P6). 
The most widespread mechanism, in lieu of the scale which occurs as a prominent theme in most images, was the act of weighing. 
A majority of the participants referenced that every theory is weighing theories against one another, hence what ends up on the scale and why it's on the scale might differ, but everybody's weighing and balancing reasons, i.e., the mode of reasoning is what differs. 

\subsubsection{Mechanisms}

The scale, depicted in Figures~\ref{fig:dbd1},~\ref{fig:dbd2},~\ref{fig:contp2},~\ref{fig:consd2}, and ~\ref{fig:pd2}, i.e. all ethical families but virtue ethics, was formalized as the ``Scales of Justice''---a classic symbol respresenting justice adopted by the legal system (P2, P6, P10). 
P8 highlighted scales to be ``endemic to how we think of justice, [and] justice endemic to how we think of ethics''. However, most participants criticized the scale to be insufficient for unanimous representation, such as in cause and effect scenarios, where it ``does not take into account all consequences'' (P3). 
To this observation, some participants highlighted networks for the holistic representation of an ethical landscape. 
Networks presented the moral landscape of an ordered universe with distinct entities and uniform links that lead to outcomes, and a procedural mechanism to trigger the relationship between interactions (P4, P6, P7). In consequential ethics, networks presented nodes as people and connections as actions that trace consequences and connections between people. In Figure~\ref{fig:consp1}, the black lines connect the cause-and-effect among people (P6), and in Figure~\ref{fig:consd1}, a council monitoring the network on the table deliberates where to ``focus attention'' (P3, P9). 

\subsection{Signs and Symbolisms}

This theme presents the signs and symbols that are used to represent ethics, where \textit{universality} and \textit{navigation} illustrate the complexity and interconnectedness of ethical theories.

\subsubsection{Universality} Universal signs and symbolisms were prevalent in the scale that is central and most fluid to ethical thinking (P8, P10). Working towards equilibrium was also perceived to be ``universal in achieving balance in ethical theories'' (P4). Following suit were the tree and cosmos for their imagery of the divine and religion through relations to telos (ultimate purpose) and morality (P1, P4, P5). The image of a light bulb was also noticed across three ethical families--- Figures~\ref{fig:vd2} (virtue ethics),~\ref{fig:contd2} (contractual ethics), and~\ref{fig:pp2} (pluralistic ethics)---as a symbol of ideation and inspiration (P4, P5, P7). 
More so, symbols in this theme were also characterized as natural and mechanical. In natural representation were the flying doves in contractual cities that signified peace and safety in Figure~\ref{fig:contp1} (P5, P6), and the idealized, virtuous imagery of the angels and fairies in Figure~\ref{fig:vd1} (P4, P6). 
In mechanical imagery, 
depictions were highlighted as items ``responding to the material foundation of idea production'' (P5), such as the presence of Lady Justice (goddess of natural and divine law, order, and justice) in Figures~\ref{fig:dbd1} (duty-based ethics) and ~\ref{fig:contp2} (contractual ethics).

\subsubsection{Navigation} Symbols of navigation were seen as ways of finding ethical direction. In Figure~\ref{fig:dbd1} (duty-based ethics), two ship wheels ``navigate their way through different duties'' (P9) and in Figure~\ref{fig:pp1} (pluralistic ethics) the compass provides the right way to go through the maze (P5, P6). In the context of pluralism, the path of discovery was described as the process of ``coming through a maze to find that theory'' (P10). A notable element of navigation surfaced by most participants was the puzzle pieces in Figure~\ref{fig:pd1} (pluralistic ethics).
Participants noticed the ``plurality of values'' (P9) in that ``no one piece is enough to complete the ethical landscape'' (P2). In the burst puzzle (Figure~\ref{fig:pd1}), P4 highlighted that one ``doesn't immediately see how they fit together [as] our views may be limited on our own''.
In some observations, the same image was alternatively read with the tree as part of the puzzle in a ``push against human-centered ethics'' (P6) where ``each tradition arises with its own soil'' (P2).  


\subsection{Image Composition}

This theme elaborates how image composition, specifically \textit{geometry} (via spatiality and centrality), and \textit{color}, represent and are used by participants to evaluate the visual cues of ethical theories.

\subsubsection{Geometry}

In the spatial composition of an image (\textit{spatiality}), participants analyzed the composition of the image as an indicator of the ethical theory. 
In Figure~\ref{fig:dbp2} (duty-based ethics), the city center, being rigid and structured, brought about ``how people think of duty-based ethics'' and the clarity associated with choice and duty (P3). The symmetry of the image was also seen as a representation of form and structure where ``no one can choose their place'' (P5), i.e., the lack of choice but duty. In pluralism, the free-flowing composition of Figure~\ref{fig:pp2} represented one that was fluid and fitting to the ethical family (P4). Likewise, in the centralized composition of an image (\textit{centrality}), participants highlighted the layout of the image as a representation of the theory. In duty-based ethics (Figures~\ref{fig:dbd1},~\ref{fig:dbd2},~\ref{fig:dbp1}), circular layouts were seen as closed-circuit systems to how the duty-based theories function (P5). In pluralism, P4 highlighted how the centrality of composition provided a lens through which there is ``no right or wrong'' as it is a theory that is unifying and simultaneous, with the spiral around the tree in Figure~\ref{fig:pd2} (pluralistic ethics) denoting a circular movement (P3). 

\subsubsection{Color}

Color served as a visual element through which an image was evaluated as an adequate representation of the ethical theory. In pluralism, the rainbow tapestry in Figure~\ref{fig:pp2} was seen to be representative of different identities, whereas the variety of colors in Figure~\ref{fig:pp2} illuminated the multicultural notion of the theory (P2, P4, P7). In the structure of societies (noted dystopia and utopia by P5), the single-tone representation in Figure~\ref{fig:dbp2} (duty-based ethics) signaled order of duty-based ethics whereas the multicolored representation in Figure~\ref{fig:contp1} (contractual ethics) was representative of what is less structured to a contractual theory (P3). P6 however highlighted how most AI imagery tends to be blue with P9 listing it to be a ``spacey background'' that is innate to AI generation. A majority of participants noted the pairing of orange and blue to be evident throughout the image compositions as well. 
This garnered scrutiny as ethical theories are multifaceted where the tone of blue was ``homogenizing'' to their distinction, i.e., in the representation of utopian and dystopian worlds, ``dystopia would have a different color scheme'' (P6). 

\section{Discussion}

Our discussion begins with reflections on the value of the kaleidoscope as a metaphor in framing our research. This is followed by 
a critical examination of the themes in our results, and cautions and considerations for GenAI models. We conclude by urging social value alignment in T2I models and describing the limitations of our work.


\subsection{Value of the Kaleidoscope Metaphor}

The use of a kaleidoscope to initiate interviews was essential to prompt deep thought into the ever-changing nature of ethical theories and GenAI models. Participants were excited to think creatively about the research question and engage in our work, especially as we presented the kaleidoscope design probe in the interviews. The format helped frame the discussion as a critical turning point that interrogated each participant's understanding of ethical theories and GenAI models. 
Together, it helped \textit{build a bridge} between ethics, a field of knowledge the participants had expertise in, and GenAI models, a field we were introducing as one that exhibited similar characteristics. The metaphor helped reduce the barrier of entry in introducing GenAI models to the experts and heightened their understanding of our work, as well as the motivation behind our research question. In summary, the kaleidoscope helped open up participants' imagination of the research question, seed the novel perspective of our research, and engrain deeper comprehension of our interdisciplinary research topic.

As a result, in one observation, the metaphor unraveled a perspective among our participants that GenAI models and ethical theories' are \textit{seemingly static yet dynamic}. Participants provided careful consideration of the generated imagery, citing observations they had anticipated or were not convinced by. This led to in-depth discussions and prompted the experts to be critical of the seemingly static phenomenon. In Science and Technology Studies (STS) literature, this observation is echoed by the ``Formalism'' and ``Ripple Effect'' traps that traditional Machine Learning systems fall into~\cite{selbst_fairness_2019}. The Formalism Trap entails a failure to account for the full meaning of social concepts, i.e. those that cannot be resolved through mathematical formalisms, and the Ripple Effect Trap entails the failure to understand how the insertion of technology into an existing social system changes the behaviors and embedded values of the pre-existing system. Our participants questioned the algorithmic coordination of judgment in the formalization of ethical concepts (Formalism Trap) and GenAI models' translation of the world from data to inherent values (Ripple Effect Trap). These observations elicit a broader focus on AI research to re-envision technical design as a process rather than a solution and engage in Value Sensitive Design (VSD)~\cite{friedman_value_2019}. 

In this regard, the unraveling of these STS traps through literary and visual art allowed our work to be a critique, an art form in itself, and a way in which the creative outlook impacted the cognition and mindset of the participants who are questioning the status quo. We take the formative role of art not only to have been the vessel of the study but also to have chiseled the participants' insight and awareness of the nature of these systems. Moreover, the use of the kaleidoscope (metaphor and design probe) offers others---researchers, practitioners, and creatives---a lens to implement art as a method to critically engage participants in research studies. The metaphor reinforces art as a method of inquiry, as elucidated in similar studies for its ability to engage audiences in critical investigation, novel perspectives, and edifying experiences ~\cite{hemment_ai_2023}. As artistic research is a growing field of inquiry, along with research-based art practices, this metaphor offers a novel way to immerse artistic inquiry into the academic and research ecosystem, especially with the growing need for interdisciplinary research~\cite{vladova_why_2024}. 

\subsection{Critical Examination of Themes}

Participants identified overlapping ideas and elements based on their understanding of the ethical families (virtue, duty-based, contractualism, consequentialism, and pluralism). This enabled the isolation of coherent themes as the basis of the \textit{Kaleidoscope Gallery}. For example, the theme of \textit{Signs and Symbolisms} formed the basis for deconstructing an image, which led to identifying the closely associated theme of \textit{Image Composition}. In particular, symbols were used by the participants to identify the inner workings of the T2I model (RQ2). They questioned the context of the model and made judgments about its implications within the ethical families that were examined. In this regard, their observations highlight context-sensitivity in ethical analysis as a meta-ethical consideration (questions about ethics itself rather than specific ethical theories)~\cite{musschenga_empirical_2005}.

Additionally, the subsequent viewing and interpretation of the generated images presented a unique platform to dissect a T2I model's visual representation of ethics. An image can be read differently by a participant if presented in one ethical family than another, prompting the sensitivity to context-framing as an integral lens for interpreting ethical theories through text-to-image relations~\cite{mitchell_picture_theory_1995}. This conveyed how each participant's understanding of GenAI models and ethics is reflected in the experiences of their counterparts. In doing so, the visualization of ethical theories garnered imagery that was diversely interpreted in their function and construction. 

Another observation indicated variability in the interpretations of the ethical families. The themes helped explain and interpret depictions of ethical families, however, themes in and of themselves, did not belong to an ethical family. In moral theory, examples of what determines the appeal of an ethical theory are its explanatory power (reasoning for decision-making), internal support (moral grounding), and determinacy (ability to provide clear and systematic answers to moral questions)~\cite[Chapter~6]{timmons_moral_2012}. Thus, an ethical family's appeal to a theme was a result of its consensual and subjective understanding among participants, and its ability to reflect inherent ability. For instance, \textit{Navigating Moral Compass} was prominent in Virtue Ethics as it situated the path of self-cultivation and provided internal support to the moral theory. In some cases, however, themes are insufficient to explain an entire family, and in such cases, participants (unprompted) chose one image from the entire exhibit as the ``best representation'' of the ethical family (RQ3). 

Lastly, participants equated their ability to analyze the imagery through their understanding of the ethical family. Their focus on themes such as society, order, law, humanity, and morality stemmed from the influence of theory within their expertise~\cite{johnson_metaphors_2003}. This represents their form of subjectivity as a curious case for \textit{context vs. bias}, possibly influenced by other factors beyond their disciplinary expertise. Gestalt principles were useful for completing the understanding of the whole image~\cite{koffka_perception_2014}, as well as color theory, where complementary colors that sit opposite each other on the color wheel (orange and blue in our gallery) created greater contrast and visual impact when used together~\cite{color_theory}. However, there still remained misinterpretation and misunderstanding due to each expert's inherent bias that influences how the image is evaluated.
Thus, while theoretical frameworks and visual principles provide valuable tools for analysis, subjective experiences remained a powerful influence on image analysis. 

\subsection{Cautions and Considerations} 

Ethics is an evolving field of critical thinking that values different perspectives to express moral truth, whereas an AI model is not capable of critically thinking and understanding its own generations. This leads to incompatible expressions within image generations in what is noted as the \textit{GenAI Paradox}. ~\citet{west2023generative_understand} explain this paradox via the inherent incapability of humans to absorb vast quantities of data. 
Meanwhile, an AI model, which is modeled over a wide set of data for reconstructing information, generates ``understanding'' that is not comprehensive to the output. We highlight the following set of cautions and considerations given this paradox.

First, we found that T2I models were able to conceptualize complex concepts, as highlighted in our themes, however, as determined by our experts, showed to be hierarchical in social construct, western in worldview, and biased in gender and geography (RQ1). Although we see their possibility of conceptual representation, there are gaps in their knowledge that suggest caution in their deployment. As socio-technical systems and not purely technical systems~\cite{kudina_sociotechnical_2024}, this brings us to the topic of social value alignment in GenAI models~\cite{ammanabrolu_aligning_2022}. ~\citet{fricker_epistemic_2024} coins the term \textit{Epistemic Injustice} to recognize the ethics relating to knowledge. The author divides this phenomenon into testimonial injustice (unfairness due to a lack of trust) and hermeneutical injustice (misunderstanding due to lack of interpretive resources )~\cite{fricker_epistemic_2024}. Recognizing epistemic injustice in T2I models, we urge models to align with artists' opinions on transparency, ownership, and fairness of AI-generated art~\cite{lovato_foregrounding_2024}. This could take shape as a community-centered participatory T2I design~\cite{ghosh_caste_2024}, whereas, in a legal context, the lack of socio-technical assessment in these models can be flagged as ``premature model deployment'' where technical development must adequately be coupled with social alignment in order to fulfill the development lifecycle. The creative community could also develop regulatory norms specific to the use of AI in visual art curation~\cite{srinivasan_see_2024} and 
engage in human-AI partnership that maximizes both partners' creative strengths~\cite{mazzone_art_2019}. 

Secondly, aligned with the metaphor of the kaleidoscope, GenAI models change based on technical alterations~\cite{lian_llm-grounded_2023}, the curation of data~\cite{rojas_dollar_nodate}, and innately by ``data'' as an amalgamation that is constant, yet fluid altogether. With training mechanisms such as AI Alignment~\cite{kenton_alignment_2021}, where prompts are used for improving a model~\cite{ouyang_training_2022}, the modality is not only altering a model's behavior but its foundational utility as the learned element of a model. We recognize the variety of open problems in this methodology~\cite{gemini,casper_open_2023, buyl_large_2024}, with specific implications for its deployment in decision-making processes~\cite{bignotti_legal_2024}, and the principles that are underway~\cite{gabriel_artificial_2020, varshney_decolonial_2024}. We present the case of ``aligning'' concepts, such as the case of subjectivity in our study, where certain perspectives are more prevalent than others, which in this methodology is embedded into the model. Thus, the training mechanisms of T2I models are effectively shifting and dynamically dependent, cautioning consideration for their design on the knowledge that is elicited through the training of perspectives. ~\citet{wallach_moral_2020} propose a shift from value alignment to virtue embodiment as a means of preserving moral codes in the engineering of AI systems.  

Lastly, AI-generated images are widely being adopted to visualize complex concepts (RQ2), where, in scientific research, have been retracted from publications and banned from certain journals due to fake data and inaccurate scientific imagery~\cite{wong_ai-generated_2024}.
The impact of this phenomenon increases with the integration of AI into pedagogy, where it serves as a teaching model for the cognitive growth of children (cautioned NSFW by our participants). In a symposium curated by the International Society of Learning Sciences~\cite{morales-navarro_making_nodate}, perspectives among scholars open perspectives on children's awareness of their world to critically understand machine learning. The pedagogy of utilizing these models as learned representations~\cite{biotechnology, math, physics, social_science} is one we present with a critical eye for perspectives of inquiry and exposition into literacy~\cite{ng_conceptualizing_2021}. Recent work has also presented 
formal taxonomies of harm to mitigate potential harms caused by AI systems~\cite{kennedy_vernacularizing_2024}. In education's unique vernacular, this entails information that appears factually correct but is verifiably incorrect (inaccurate) or reflective of reasoning that has been rejected (inauthentic). One approach to this disparity has been cultivating a socio-cultural understanding of AI in children through participatory design~\cite{dangol_mediating_2024}.

\subsection{Limitations and Future Work}

In this section, we address the ethics of using GenAI models for this work. This includes our use of T2I models and the exploitation of artists' work in these models~\cite{goetze_theft_2024}. We also acknowledge the biases that are likely present in DALL-E 3 (the T2I model). As such, this work is a visual inquiry of ethical theories and thus does not warrant a direct representation of the ethical theories by DALL-E 3. More so, being situated in a Western institution skewed the ethical theories to be primarily from Western philosophy (e.g. lacking the African Ubuntu), which we acknowledge as bias perpetuated from the study and work that will require engagement from participants across the globe.

We also acknowledge the biases the participants might have towards GenAI that might impact their participation, and the biases, we as researchers, might have perpetuated in the images created, such as the curation of \textit{Kaleidoscope Gallery} with images corresponding to an ethical family. The latter was conducted to provide context to the interpretation of the images although its potential influence on the experts' interpretations of images is one we present for future work. 

Lastly, we assess the impact of generating four images per ethical family. We acknowledge that this small sample might limit potential arguments about inherent bias within the models and the generalizability of our findings towards a complete kaleidoscope of how ethics is depicted by GenAI. However, we made this deliberate design choice not to overcrowd the experts' evaluation of images and to provide a detailed scrutiny for each image. In future work, we aim to curate images over multiple periods to generate a time series of T2I images
and to address the limited sample size of our curation. We also plan to include other T2I models (e.g., MidJourney, Stable Diffusion) to diversify our study. 

\section{Conclusion}

In this paper, we investigate ethical theories and GenAI models through art and the emerging field of \textit{Visual Ethics} and carry out a qualitative analysis to understand the underlying assumptions of GenAI models. Through the curation of \textit{Kaleidoscope Gallery} via T2I models, our findings indicate morality, society and learned associations to be central tenets in the moral imagination of ethics. Our investigation highlights the diversity and fluidity of ethical theories, and the contextual and biased impact of worldview expression on model behavior. We present this work to critically examine foundational knowledge in GenAI models and provide cautions and considerations for knowledge dissemination, model development, and pedagogy. We call for a sociologically robust approach to 
social value alignment in T2I models.

\begin{acks}
We would like to thank our interview participants for sharing their insights, wisdom, and knowledge. 
\end{acks}

\bibliographystyle{ACM-Reference-Format}
\bibliography{biblio}


\begin{thebibliography}{97}


\ifx \showCODEN    \undefined \def \showCODEN     #1{\unskip}     \fi
\ifx \showDOI      \undefined \def \showDOI       #1{#1}\fi
\ifx \showISBNx    \undefined \def \showISBNx     #1{\unskip}     \fi
\ifx \showISBNxiii \undefined \def \showISBNxiii  #1{\unskip}     \fi
\ifx \showISSN     \undefined \def \showISSN      #1{\unskip}     \fi
\ifx \showLCCN     \undefined \def \showLCCN      #1{\unskip}     \fi
\ifx \shownote     \undefined \def \shownote      #1{#1}          \fi
\ifx \showarticletitle \undefined \def \showarticletitle #1{#1}   \fi
\ifx \showURL      \undefined \def \showURL       {\relax}        \fi
\providecommand\bibfield[2]{#2}
\providecommand\bibinfo[2]{#2}
\providecommand\natexlab[1]{#1}
\providecommand\showeprint[2][]{arXiv:#2}

\bibitem[Aizenberg and van~den Hoven(2020)]%
        {human_rights_in_AI}
\bibfield{author}{\bibinfo{person}{Evgeni Aizenberg} {and} \bibinfo{person}{Jeroen van~den Hoven}.} \bibinfo{year}{2020}\natexlab{}.
\newblock \bibinfo{title}{Designing for human rights in {AI}}.
\newblock
\newblock
\urldef\tempurl%
\url{https://doi.org/10.1177/2053951720949566}
\showDOI{\tempurl}


\bibitem[Alemohammad et~al\mbox{.}(2023)]%
        {alemohammad_self-consuming_2023}
\bibfield{author}{\bibinfo{person}{Sina Alemohammad}, \bibinfo{person}{Josue Casco-Rodriguez}, \bibinfo{person}{Lorenzo Luzi}, \bibinfo{person}{Ahmed~Imtiaz Humayun}, \bibinfo{person}{Hossein Babaei}, \bibinfo{person}{Daniel LeJeune}, \bibinfo{person}{Ali Siahkoohi}, {and} \bibinfo{person}{Richard~G. Baraniuk}.} \bibinfo{year}{2023}\natexlab{}.
\newblock \bibinfo{title}{Self-{Consuming} {Generative} {Models} {Go} {MAD}}.
\newblock
\newblock
\urldef\tempurl%
\url{https://doi.org/10.48550/arXiv.2307.01850}
\showDOI{\tempurl}
\newblock
\shownote{arXiv:2307.01850}.


\bibitem[Alexander and Moore(2024)]%
        {deontological_alexander_2024}
\bibfield{author}{\bibinfo{person}{Larry Alexander} {and} \bibinfo{person}{Michael Moore}.} \bibinfo{year}{2024}\natexlab{}.
\newblock \showarticletitle{Deontological {Ethics}}.
\newblock In \bibinfo{booktitle}{\emph{The {Stanford} {Encyclopedia} of {Philosophy}} (\bibinfo{edition}{winter 2024} ed.)}, \bibfield{editor}{\bibinfo{person}{Edward~N. Zalta} {and} \bibinfo{person}{Uri Nodelman}} (Eds.). \bibinfo{publisher}{Metaphysics Research Lab, Stanford University}.
\newblock
\urldef\tempurl%
\url{https://plato.stanford.edu/archives/win2024/entries/ethics-deontological/}
\showURL{%
\tempurl}


\bibitem[Ammanabrolu et~al\mbox{.}(2022)]%
        {ammanabrolu_aligning_2022}
\bibfield{author}{\bibinfo{person}{Prithviraj Ammanabrolu}, \bibinfo{person}{Liwei Jiang}, \bibinfo{person}{Maarten Sap}, \bibinfo{person}{Hannaneh Hajishirzi}, {and} \bibinfo{person}{Yejin Choi}.} \bibinfo{year}{2022}\natexlab{}.
\newblock \bibinfo{title}{Aligning to {Social} {Norms} and {Values} in {Interactive} {Narratives}}.
\newblock
\newblock
\urldef\tempurl%
\url{https://doi.org/10.48550/arXiv.2205.01975}
\showDOI{\tempurl}
\newblock
\shownote{arXiv:2205.01975}.


\bibitem[Ashford and Mulgan(2018)]%
        {contractualism_ashford_2018}
\bibfield{author}{\bibinfo{person}{Elizabeth Ashford} {and} \bibinfo{person}{Tim Mulgan}.} \bibinfo{year}{2018}\natexlab{}.
\newblock \showarticletitle{Contractualism}.
\newblock In \bibinfo{booktitle}{\emph{The {Stanford} {Encyclopedia} of {Philosophy}} (\bibinfo{edition}{summer 2018} ed.)}, \bibfield{editor}{\bibinfo{person}{Edward~N. Zalta}} (Ed.). \bibinfo{publisher}{Metaphysics Research Lab, Stanford University}.
\newblock
\urldef\tempurl%
\url{https://plato.stanford.edu/archives/sum2018/entries/contractualism/}
\showURL{%
\tempurl}


\bibitem[Bennett et~al\mbox{.}(2013)]%
        {bennett_agricultural_2013}
\bibfield{author}{\bibinfo{person}{Alan~B. Bennett}, \bibinfo{person}{Cecilia Chi-Ham}, \bibinfo{person}{Geoffrey Barrows}, \bibinfo{person}{Steven Sexton}, {and} \bibinfo{person}{David Zilberman}.} \bibinfo{year}{2013}\natexlab{}.
\newblock \showarticletitle{Agricultural {Biotechnology}: {Economics}, {Environment}, {Ethics}, and the {Future}}.
\newblock \bibinfo{journal}{\emph{Annual Review of Environment and Resources}} \bibinfo{volume}{38}, \bibinfo{number}{1} (\bibinfo{year}{2013}), \bibinfo{pages}{249--279}.
\newblock
\urldef\tempurl%
\url{https://doi.org/10.1146/annurev-environ-050912-124612}
\showDOI{\tempurl}
\newblock
\shownote{\_eprint: https://doi.org/10.1146/annurev-environ-050912-124612}.


\bibitem[Betker et~al\mbox{.}(2023)]%
        {betker_improving_dalle3}
\bibfield{author}{\bibinfo{person}{James Betker}, \bibinfo{person}{Gabriel Goh}, \bibinfo{person}{Li Jing}, \bibinfo{person}{Tim Brooks}, \bibinfo{person}{Jianfeng Wang}, \bibinfo{person}{Linjie Li}, \bibinfo{person}{Long Ouyang}, \bibinfo{person}{Juntang Zhuang}, \bibinfo{person}{Joyce Lee}, \bibinfo{person}{Yufei Guo}, \bibinfo{person}{Wesam Manassra}, \bibinfo{person}{Prafulla Dhariwal}, \bibinfo{person}{Casey Chu}, \bibinfo{person}{Yunxin Jiao}, {and} \bibinfo{person}{Aditya Ramesh}.} \bibinfo{year}{2023}\natexlab{}.
\newblock \showarticletitle{Improving {Image} {Generation} with {Better} {Captions}}.
\newblock  (\bibinfo{year}{2023}).
\newblock


\bibitem[Bignotti and Camassa(2024)]%
        {bignotti_legal_2024}
\bibfield{author}{\bibinfo{person}{Camilla Bignotti} {and} \bibinfo{person}{Carolina Camassa}.} \bibinfo{year}{2024}\natexlab{}.
\newblock \bibinfo{title}{Legal {Minds}, {Algorithmic} {Decisions}: {How} {LLMs} {Apply} {Constitutional} {Principles} in {Complex} {Scenarios}}.
\newblock
\newblock
\urldef\tempurl%
\url{https://doi.org/10.48550/arXiv.2407.19760}
\showDOI{\tempurl}
\newblock
\shownote{arXiv:2407.19760}.


\bibitem[Bird et~al\mbox{.}(2023)]%
        {bird_typology_2023}
\bibfield{author}{\bibinfo{person}{Charlotte Bird}, \bibinfo{person}{Eddie Ungless}, {and} \bibinfo{person}{Atoosa Kasirzadeh}.} \bibinfo{year}{2023}\natexlab{}.
\newblock \showarticletitle{Typology of {Risks} of {Generative} {Text}-to-{Image} {Models}}. In \bibinfo{booktitle}{\emph{Proceedings of the 2023 {AAAI}/{ACM} {Conference} on {AI}, {Ethics}, and {Society}}}. \bibinfo{publisher}{ACM}, \bibinfo{address}{Montr{\textbackslash}'\{e\}al QC Canada}, \bibinfo{pages}{396--410}.
\newblock
\showISBNx{9798400702310}
\urldef\tempurl%
\url{https://doi.org/10.1145/3600211.3604722}
\showDOI{\tempurl}


\bibitem[Buyl et~al\mbox{.}(2024)]%
        {buyl_large_2024}
\bibfield{author}{\bibinfo{person}{Maarten Buyl}, \bibinfo{person}{Alexander Rogiers}, \bibinfo{person}{Sander Noels}, \bibinfo{person}{Iris Dominguez-Catena}, \bibinfo{person}{Edith Heiter}, \bibinfo{person}{Raphael Romero}, \bibinfo{person}{Iman Johary}, \bibinfo{person}{Alexandru-Cristian Mara}, \bibinfo{person}{Jefrey Lijffijt}, {and} \bibinfo{person}{Tijl~De Bie}.} \bibinfo{year}{2024}\natexlab{}.
\newblock \bibinfo{title}{Large {Language} {Models} {Reflect} the {Ideology} of their {Creators}}.
\newblock
\newblock
\urldef\tempurl%
\url{http://arxiv.org/abs/2410.18417}
\showURL{%
\tempurl}
\newblock
\shownote{arXiv:2410.18417}.


\bibitem[Casper et~al\mbox{.}(2023)]%
        {casper_open_2023}
\bibfield{author}{\bibinfo{person}{Stephen Casper}, \bibinfo{person}{Xander Davies}, \bibinfo{person}{Claudia Shi}, \bibinfo{person}{Thomas~Krendl Gilbert}, \bibinfo{person}{Jérémy Scheurer}, \bibinfo{person}{Javier Rando}, \bibinfo{person}{Rachel Freedman}, \bibinfo{person}{Tomasz Korbak}, \bibinfo{person}{David Lindner}, \bibinfo{person}{Pedro Freire}, \bibinfo{person}{Tony Wang}, \bibinfo{person}{Samuel Marks}, \bibinfo{person}{Charbel-Raphaël Segerie}, \bibinfo{person}{Micah Carroll}, \bibinfo{person}{Andi Peng}, \bibinfo{person}{Phillip Christoffersen}, \bibinfo{person}{Mehul Damani}, \bibinfo{person}{Stewart Slocum}, \bibinfo{person}{Usman Anwar}, \bibinfo{person}{Anand Siththaranjan}, \bibinfo{person}{Max Nadeau}, \bibinfo{person}{Eric~J. Michaud}, \bibinfo{person}{Jacob Pfau}, \bibinfo{person}{Dmitrii Krasheninnikov}, \bibinfo{person}{Xin Chen}, \bibinfo{person}{Lauro Langosco}, \bibinfo{person}{Peter Hase}, \bibinfo{person}{Erdem Bıyık}, \bibinfo{person}{Anca Dragan}, \bibinfo{person}{David
  Krueger}, \bibinfo{person}{Dorsa Sadigh}, {and} \bibinfo{person}{Dylan Hadfield-Menell}.} \bibinfo{year}{2023}\natexlab{}.
\newblock \bibinfo{title}{Open {Problems} and {Fundamental} {Limitations} of {Reinforcement} {Learning} from {Human} {Feedback}}.
\newblock
\newblock
\urldef\tempurl%
\url{http://arxiv.org/abs/2307.15217}
\showURL{%
\tempurl}
\newblock
\shownote{arXiv:2307.15217 [cs]}.


\bibitem[Cetinic and She(2021)]%
        {cetinic_understanding_2021}
\bibfield{author}{\bibinfo{person}{Eva Cetinic} {and} \bibinfo{person}{James She}.} \bibinfo{year}{2021}\natexlab{}.
\newblock \bibinfo{title}{Understanding and {Creating} {Art} with {AI}: {Review} and {Outlook}}.
\newblock
\newblock
\urldef\tempurl%
\url{https://doi.org/10.48550/arXiv.2102.09109}
\showDOI{\tempurl}
\newblock
\shownote{arXiv:2102.09109}.


\bibitem[Dangol et~al\mbox{.}(2024)]%
        {dangol_mediating_2024}
\bibfield{author}{\bibinfo{person}{Aayushi Dangol}, \bibinfo{person}{Michele Newman}, \bibinfo{person}{Robert Wolfe}, \bibinfo{person}{Jin~Ha Lee}, \bibinfo{person}{Julie~A. Kientz}, \bibinfo{person}{Jason Yip}, {and} \bibinfo{person}{Caroline Pitt}.} \bibinfo{year}{2024}\natexlab{}.
\newblock \showarticletitle{Mediating {Culture}: {Cultivating} {Socio}-cultural {Understanding} of {AI} in {Children} through {Participatory} {Design}}. In \bibinfo{booktitle}{\emph{Designing {Interactive} {Systems} {Conference}}}. \bibinfo{publisher}{ACM}, \bibinfo{address}{IT University of Copenhagen Denmark}, \bibinfo{pages}{1805--1822}.
\newblock
\showISBNx{9798400705830}
\urldef\tempurl%
\url{https://doi.org/10.1145/3643834.3661515}
\showDOI{\tempurl}


\bibitem[Daniele and Song(2019)]%
        {daniele_ai_2019}
\bibfield{author}{\bibinfo{person}{Antonio Daniele} {and} \bibinfo{person}{Yi-Zhe Song}.} \bibinfo{year}{2019}\natexlab{}.
\newblock \showarticletitle{{AI} + {Art} = {Human}}. In \bibinfo{booktitle}{\emph{Proceedings of the 2019 {AAAI}/{ACM} {Conference} on {AI}, {Ethics}, and {Society}}}. \bibinfo{publisher}{ACM}, \bibinfo{address}{Honolulu HI USA}, \bibinfo{pages}{155--161}.
\newblock
\showISBNx{978-1-4503-6324-2}
\urldef\tempurl%
\url{https://doi.org/10.1145/3306618.3314233}
\showDOI{\tempurl}


\bibitem[Ernest(2018)]%
        {ernest_pictograms_2018}
\bibfield{author}{\bibinfo{person}{Ernest}.} \bibinfo{year}{2018}\natexlab{}.
\newblock \bibinfo{title}{Pictograms and {Ideograms} - {Origin} and {History}}.
\newblock
\newblock
\urldef\tempurl%
\url{https://citrinitas.com/a-history-of-visual-communication-pictograms-and-ideograms/}
\showURL{%
\tempurl}


\bibitem[{Face}(2024)]%
        {hf_t2i_2024}
\bibfield{author}{\bibinfo{person}{{Hugging} {Face}}.} \bibinfo{year}{2024}\natexlab{}.
\newblock \bibinfo{title}{What is {Text}-to-{Image}?}
\newblock
\newblock
\urldef\tempurl%
\url{https://huggingface.co/tasks/text-to-image}
\showURL{%
\tempurl}


\bibitem[Flis and Piotrowski(2024)]%
        {flis_conceptual_metaphor_2024}
\bibfield{author}{\bibinfo{person}{Maria Flis} {and} \bibinfo{person}{Karol Piotrowski}.} \bibinfo{year}{2024}\natexlab{}.
\newblock \showarticletitle{The {Conceptual} {Metaphor} as an {Ethical} {Kaleidoscope} in {Field} {Research}}.
\newblock \bibinfo{journal}{\emph{Qualitative Sociology Review}} \bibinfo{volume}{20}, \bibinfo{number}{1} (\bibinfo{date}{Jan.} \bibinfo{year}{2024}), \bibinfo{pages}{30--41}.
\newblock
\showISSN{1733-8077}
\urldef\tempurl%
\url{https://doi.org/10.18778/1733-8077.20.1.03}
\showDOI{\tempurl}


\bibitem[Foundation(2016)]%
        {color_theory}
\bibfield{author}{\bibinfo{person}{Interaction~Design Foundation}.} \bibinfo{year}{2016}\natexlab{}.
\newblock \bibinfo{title}{What is {Color} {Theory}?}
\newblock
\newblock
\urldef\tempurl%
\url{https://www.interaction-design.org/literature/topics/color-theory}
\showURL{%
\tempurl}


\bibitem[Fricker(2024)]%
        {fricker_epistemic_2024}
\bibfield{author}{\bibinfo{person}{Miranda Fricker}.} \bibinfo{year}{2024}\natexlab{}.
\newblock \bibinfo{title}{Epistemic injustice}.
\newblock
\newblock
\urldef\tempurl%
\url{https://en.wikipedia.org/w/index.php?title=Epistemic_injustice&oldid=1251981459}
\showURL{%
\tempurl}
\newblock
\shownote{Page Version ID: 1251981459}.


\bibitem[Friedman and Hendry(2019)]%
        {friedman_value_2019}
\bibfield{author}{\bibinfo{person}{Batya Friedman} {and} \bibinfo{person}{David~G. Hendry}.} \bibinfo{year}{2019}\natexlab{}.
\newblock \bibinfo{booktitle}{\emph{Value Sensitive Design: Shaping Technology with Moral Imagination}}.
\newblock \bibinfo{publisher}{The MIT Press}.
\newblock
\showISBNx{9780262039536}
\urldef\tempurl%
\url{https://mitpress.mit.edu/9780262039536/value-sensitive-design/}
\showURL{%
\tempurl}


\bibitem[Gabriel(2020)]%
        {gabriel_artificial_2020}
\bibfield{author}{\bibinfo{person}{Iason Gabriel}.} \bibinfo{year}{2020}\natexlab{}.
\newblock \showarticletitle{Artificial {Intelligence}, {Values} and {Alignment}}.
\newblock \bibinfo{journal}{\emph{Minds and Machines}} \bibinfo{volume}{30}, \bibinfo{number}{3} (\bibinfo{date}{Sept.} \bibinfo{year}{2020}), \bibinfo{pages}{411--437}.
\newblock
\showISSN{0924-6495, 1572-8641}
\urldef\tempurl%
\url{https://doi.org/10.1007/s11023-020-09539-2}
\showDOI{\tempurl}
\newblock
\shownote{arXiv:2001.09768 [cs]}.


\bibitem[Gainetdinov(2023)]%
        {gainetdinov_diffusion_2023}
\bibfield{author}{\bibinfo{person}{Ainur Gainetdinov}.} \bibinfo{year}{2023}\natexlab{}.
\newblock \bibinfo{title}{Diffusion {Models} vs {GANs} vs {VAEs}: {Comparison} of {Deep} {Generative} {Models}}.
\newblock
\newblock
\urldef\tempurl%
\url{https://pub.towardsai.net/diffusion-models-vs-gans-vs-vaes-comparison-of-deep-generative-models-67ab93e0d9ae}
\showURL{%
\tempurl}


\bibitem[Gefen et~al\mbox{.}(2021)]%
        {social_science}
\bibfield{author}{\bibinfo{person}{Alexandre Gefen}, \bibinfo{person}{Léa Saint-Raymond}, {and} \bibinfo{person}{Tommaso Venturini}.} \bibinfo{year}{2021}\natexlab{}.
\newblock \showarticletitle{AI for Digital Humanities and Computational Social Sciences}.
\newblock \bibinfo{journal}{\emph{Lecture Notes in Computer Science}} (\bibinfo{year}{2021}).
\newblock
\urldef\tempurl%
\url{https://doi.org/10.1007/978-3-030-69128-8_12}
\showDOI{\tempurl}


\bibitem[Ghosh(2024)]%
        {ghosh_caste_2024}
\bibfield{author}{\bibinfo{person}{Sourojit Ghosh}.} \bibinfo{year}{2024}\natexlab{}.
\newblock \bibinfo{title}{Interpretations, {Representations}, and {Stereotypes} of {Caste} within {Text}-to-{Image} {Generators}}.
\newblock
\newblock
\urldef\tempurl%
\url{https://arxiv.org/html/2408.01590v1}
\showURL{%
\tempurl}


\bibitem[Goetze(2024)]%
        {goetze_theft_2024}
\bibfield{author}{\bibinfo{person}{Trystan~S. Goetze}.} \bibinfo{year}{2024}\natexlab{}.
\newblock \showarticletitle{Ai Art is Theft: Labour, Extraction, and Exploitation, or, on the Dangers of Stochastic Pollocks}.
\newblock \bibinfo{journal}{\emph{Proceedings of the 2024 Acm Conference on Fairness, Accountability, and Transparency}} (\bibinfo{year}{2024}), \bibinfo{pages}{186--196}.
\newblock


\bibitem[Goh and Sze(2019)]%
        {biotechnology}
\bibfield{author}{\bibinfo{person}{Wilson Wen~Bin Goh} {and} \bibinfo{person}{Chun~Chau Sze}.} \bibinfo{year}{2019}\natexlab{}.
\newblock \showarticletitle{AI Paradigms for Teaching Biotechnology}.
\newblock \bibinfo{journal}{\emph{Trends in Biotechnology}} (\bibinfo{year}{2019}).
\newblock
\urldef\tempurl%
\url{https://doi.org/10.1016/J.TIBTECH.2018.09.009}
\showDOI{\tempurl}


\bibitem[Goodman(1961)]%
        {snowball}
\bibfield{author}{\bibinfo{person}{Leo~A. Goodman}.} \bibinfo{year}{1961}\natexlab{}.
\newblock \showarticletitle{{Snowball Sampling}}.
\newblock \bibinfo{journal}{\emph{The Annals of Mathematical Statistics}} \bibinfo{volume}{32}, \bibinfo{number}{1} (\bibinfo{year}{1961}), \bibinfo{pages}{148 -- 170}.
\newblock
\urldef\tempurl%
\url{https://doi.org/10.1214/aoms/1177705148}
\showDOI{\tempurl}


\bibitem[Heilinger(2022)]%
        {heilinger_ethics_of_ethics_2022}
\bibfield{author}{\bibinfo{person}{Jan-Christoph Heilinger}.} \bibinfo{year}{2022}\natexlab{}.
\newblock \showarticletitle{The {Ethics} of {AI} {Ethics}. {A} {Constructive} {Critique}}.
\newblock \bibinfo{journal}{\emph{Philosophy \& Technology}} \bibinfo{volume}{35}, \bibinfo{number}{3} (\bibinfo{date}{July} \bibinfo{year}{2022}), \bibinfo{pages}{61}.
\newblock
\showISSN{2210-5441}
\urldef\tempurl%
\url{https://doi.org/10.1007/s13347-022-00557-9}
\showDOI{\tempurl}


\bibitem[Hemment et~al\mbox{.}(2023)]%
        {hemment_ai_2023}
\bibfield{author}{\bibinfo{person}{Drew Hemment}, \bibinfo{person}{Morgan Currie}, \bibinfo{person}{Sj Bennett}, \bibinfo{person}{Jake Elwes}, \bibinfo{person}{Anna Ridler}, \bibinfo{person}{Caroline Sinders}, \bibinfo{person}{Matjaz Vidmar}, \bibinfo{person}{Robin Hill}, {and} \bibinfo{person}{Holly Warner}.} \bibinfo{year}{2023}\natexlab{}.
\newblock \showarticletitle{{AI} in the {Public} {Eye}: {Investigating} {Public} {AI} {Literacy} {Through} {AI} {Art}}. In \bibinfo{booktitle}{\emph{2023 {ACM} {Conference} on {Fairness}, {Accountability}, and {Transparency}}}. \bibinfo{publisher}{ACM}, \bibinfo{address}{Chicago IL USA}, \bibinfo{pages}{931--942}.
\newblock
\showISBNx{9798400701924}
\urldef\tempurl%
\url{https://doi.org/10.1145/3593013.3594052}
\showDOI{\tempurl}


\bibitem[Hendrycks et~al\mbox{.}(2023)]%
        {ethics_dataset_2023}
\bibfield{author}{\bibinfo{person}{Dan Hendrycks}, \bibinfo{person}{Collin Burns}, \bibinfo{person}{Steven Basart}, \bibinfo{person}{Andrew Critch}, \bibinfo{person}{Jerry Li}, \bibinfo{person}{Dawn Song}, {and} \bibinfo{person}{Jacob Steinhardt}.} \bibinfo{year}{2023}\natexlab{}.
\newblock \bibinfo{title}{Aligning {AI} {With} {Shared} {Human} {Values}}.
\newblock
\newblock
\urldef\tempurl%
\url{http://arxiv.org/abs/2008.02275}
\showURL{%
\tempurl}
\newblock
\shownote{arXiv:2008.02275 [cs]}.


\bibitem[Hospers(1954)]%
        {art_expression}
\bibfield{author}{\bibinfo{person}{John Hospers}.} \bibinfo{year}{1954}\natexlab{}.
\newblock \showarticletitle{The {Concept} of {Artistic} {Expression}}.
\newblock \bibinfo{journal}{\emph{Proceedings of the Aristotelian Society}}  \bibinfo{volume}{55} (\bibinfo{year}{1954}), \bibinfo{pages}{313--344}.
\newblock
\showISSN{0066-7374}
\urldef\tempurl%
\url{https://www.jstor.org/stable/4544551}
\showURL{%
\tempurl}
\newblock
\shownote{Publisher: [Aristotelian Society, Wiley]}.


\bibitem[Hursthouse and Pettigrove(2023)]%
        {virtue_hursthouse_2023}
\bibfield{author}{\bibinfo{person}{Rosalind Hursthouse} {and} \bibinfo{person}{Glen Pettigrove}.} \bibinfo{year}{2023}\natexlab{}.
\newblock \showarticletitle{Virtue {Ethics}}.
\newblock In \bibinfo{booktitle}{\emph{The {Stanford} {Encyclopedia} of {Philosophy}} (\bibinfo{edition}{fall 2023} ed.)}, \bibfield{editor}{\bibinfo{person}{Edward~N. Zalta} {and} \bibinfo{person}{Uri Nodelman}} (Eds.). \bibinfo{publisher}{Metaphysics Research Lab, Stanford University}.
\newblock
\urldef\tempurl%
\url{https://plato.stanford.edu/archives/fall2023/entries/ethics-virtue/}
\showURL{%
\tempurl}


\bibitem[Iluz et~al\mbox{.}(2023)]%
        {iluz_word-as-image_2023}
\bibfield{author}{\bibinfo{person}{Shir Iluz}, \bibinfo{person}{Yael Vinker}, \bibinfo{person}{Amir Hertz}, \bibinfo{person}{Daniel Berio}, \bibinfo{person}{Daniel Cohen-Or}, {and} \bibinfo{person}{Ariel Shamir}.} \bibinfo{year}{2023}\natexlab{}.
\newblock \bibinfo{title}{Word-{As}-{Image} for {Semantic} {Typography}}.
\newblock
\newblock
\urldef\tempurl%
\url{http://arxiv.org/abs/2303.01818}
\showURL{%
\tempurl}
\newblock
\shownote{arXiv:2303.01818 [cs]}.


\bibitem[Jia et~al\mbox{.}(2023)]%
        {democratic_embedding}
\bibfield{author}{\bibinfo{person}{Chenyan Jia}, \bibinfo{person}{Michelle~S. Lam}, \bibinfo{person}{Minh~Chau Mai}, \bibinfo{person}{Jeff Hancock}, {and} \bibinfo{person}{Michael~S. Bernstein}.} \bibinfo{year}{2023}\natexlab{}.
\newblock \bibinfo{title}{Embedding Democratic Values into Social Media AIs via Societal Objective Functions}.
\newblock
\newblock
\showeprint[arxiv]{2307.13912}~[cs.HC]


\bibitem[Jiang et~al\mbox{.}(2022)]%
        {jiang_delphi}
\bibfield{author}{\bibinfo{person}{Liwei Jiang}, \bibinfo{person}{Jena~D. Hwang}, \bibinfo{person}{Chandra Bhagavatula}, \bibinfo{person}{Ronan~Le Bras}, \bibinfo{person}{Jenny Liang}, \bibinfo{person}{Jesse Dodge}, \bibinfo{person}{Keisuke Sakaguchi}, \bibinfo{person}{Maxwell Forbes}, \bibinfo{person}{Jon Borchardt}, \bibinfo{person}{Saadia Gabriel}, \bibinfo{person}{Yulia Tsvetkov}, \bibinfo{person}{Oren Etzioni}, \bibinfo{person}{Maarten Sap}, \bibinfo{person}{Regina Rini}, {and} \bibinfo{person}{Yejin Choi}.} \bibinfo{year}{2022}\natexlab{}.
\newblock \bibinfo{title}{Can {Machines} {Learn} {Morality}? {The} {Delphi} {Experiment}}.
\newblock
\newblock
\urldef\tempurl%
\url{https://doi.org/10.48550/arXiv.2110.07574}
\showDOI{\tempurl}
\newblock
\shownote{arXiv:2110.07574 [cs]}.


\bibitem[Johnson(1994)]%
        {johnson_moralimagination_1994}
\bibfield{author}{\bibinfo{person}{Mark Johnson}.} \bibinfo{year}{1994}\natexlab{}.
\newblock \bibinfo{booktitle}{\emph{Moral {Imagination}: {Implications} of {Cognitive} {Science} for {Ethics}}}.
\newblock \bibinfo{publisher}{University of Chicago Press}, \bibinfo{address}{Chicago, IL}.
\newblock
\showISBNx{978-0-226-40169-0}
\urldef\tempurl%
\url{https://press.uchicago.edu/ucp/books/book/chicago/M/bo3684141.html}
\showURL{%
\tempurl}


\bibitem[Kennedy and Campos(2024)]%
        {kennedy_vernacularizing_2024}
\bibfield{author}{\bibinfo{person}{Wm~Matthew Kennedy} {and} \bibinfo{person}{Daniel~Vargas Campos}.} \bibinfo{year}{2024}\natexlab{}.
\newblock \bibinfo{title}{Vernacularizing {Taxonomies} of {Harm} is {Essential} for {Operationalizing} {Holistic} {AI} {Safety}}.
\newblock
\newblock
\urldef\tempurl%
\url{https://doi.org/10.48550/arXiv.2410.16562}
\showDOI{\tempurl}
\newblock
\shownote{arXiv:2410.16562}.


\bibitem[Kenton et~al\mbox{.}(2021)]%
        {kenton_alignment_2021}
\bibfield{author}{\bibinfo{person}{Zachary Kenton}, \bibinfo{person}{Tom Everitt}, \bibinfo{person}{Laura Weidinger}, \bibinfo{person}{Iason Gabriel}, \bibinfo{person}{Vladimir Mikulik}, {and} \bibinfo{person}{Geoffrey Irving}.} \bibinfo{year}{2021}\natexlab{}.
\newblock \bibinfo{title}{Alignment of {Language} {Agents}}.
\newblock
\newblock
\urldef\tempurl%
\url{http://arxiv.org/abs/2103.14659}
\showURL{%
\tempurl}
\newblock
\shownote{arXiv:2103.14659 [cs]}.


\bibitem[Koffka(2014)]%
        {koffka_perception_2014}
\bibfield{author}{\bibinfo{person}{Kurt Koffka}.} \bibinfo{year}{2014}\natexlab{}.
\newblock \bibinfo{booktitle}{\emph{Perception: {An} {Introduction} {To} {The} {Gestalt} {Theory}: {A} {Classic} {Article} in the {History} of {Psychology}}}.
\newblock \bibinfo{publisher}{www.all-about-psychology.com}.
\newblock


\bibitem[Kudina and van~de Poel(2024)]%
        {kudina_sociotechnical_2024}
\bibfield{author}{\bibinfo{person}{Olya Kudina} {and} \bibinfo{person}{Ibo van~de Poel}.} \bibinfo{year}{2024}\natexlab{}.
\newblock \showarticletitle{A sociotechnical system perspective on {AI}}.
\newblock \bibinfo{journal}{\emph{Minds and Machines}} \bibinfo{volume}{34}, \bibinfo{number}{3} (\bibinfo{date}{June} \bibinfo{year}{2024}), \bibinfo{pages}{21}.
\newblock
\showISSN{1572-8641}
\urldef\tempurl%
\url{https://doi.org/10.1007/s11023-024-09680-2}
\showDOI{\tempurl}


\bibitem[Lakoff and Johnson(2003)]%
        {johnson_metaphors_2003}
\bibfield{author}{\bibinfo{person}{George Lakoff} {and} \bibinfo{person}{Mark Johnson}.} \bibinfo{year}{2003}\natexlab{}.
\newblock \bibinfo{booktitle}{\emph{Metaphors {We} {Live} {By}}}.
\newblock \bibinfo{publisher}{University of Chicago Press}, \bibinfo{address}{Chicago, IL}.
\newblock
\showISBNx{978-0-226-46801-3}
\urldef\tempurl%
\url{https://press.uchicago.edu/ucp/books/book/chicago/M/bo3637992.html}
\showURL{%
\tempurl}


\bibitem[Landecker(1999)]%
        {landecker_dna_1999}
\bibfield{author}{\bibinfo{person}{Hannah Landecker}.} \bibinfo{year}{1999}\natexlab{}.
\newblock \showarticletitle{Between {Beneficence} and {Chattel}: {The} {Human} {Biological} in {Law} and {Science}}.
\newblock \bibinfo{journal}{\emph{Science in Context}} \bibinfo{volume}{12}, \bibinfo{number}{1} (\bibinfo{date}{April} \bibinfo{year}{1999}), \bibinfo{pages}{203--225}.
\newblock
\showISSN{1474-0664, 0269-8897}
\urldef\tempurl%
\url{https://doi.org/10.1017/S0269889700003367}
\showDOI{\tempurl}
\newblock
\shownote{Publisher: Cambridge University Press}.


\bibitem[Lee et~al\mbox{.}(2022)]%
        {math}
\bibfield{author}{\bibinfo{person}{Dabae Lee}, \bibinfo{person}{Sheunghyun Yeo}, \bibinfo{person}{Dabae Lee}, {and} \bibinfo{person}{Sheunghyun Yeo}.} \bibinfo{year}{2022}\natexlab{}.
\newblock \showarticletitle{Developing an AI-based chatbot for practicing responsive teaching in mathematics}.
\newblock  (\bibinfo{year}{2022}).
\newblock
\urldef\tempurl%
\url{https://doi.org/10.1016/J.COMPEDU.2022.104646}
\showDOI{\tempurl}


\bibitem[Lee et~al\mbox{.}(2023)]%
        {lee_rlaif_2023}
\bibfield{author}{\bibinfo{person}{Harrison Lee}, \bibinfo{person}{Samrat Phatale}, \bibinfo{person}{Hassan Mansoor}, \bibinfo{person}{Kellie Lu}, \bibinfo{person}{Thomas Mesnard}, \bibinfo{person}{Colton Bishop}, \bibinfo{person}{Victor Carbune}, {and} \bibinfo{person}{Abhinav Rastogi}.} \bibinfo{year}{2023}\natexlab{}.
\newblock \bibinfo{title}{{RLAIF}: {Scaling} {Reinforcement} {Learning} from {Human} {Feedback} with {AI} {Feedback}}.
\newblock
\newblock
\urldef\tempurl%
\url{http://arxiv.org/abs/2309.00267}
\showURL{%
\tempurl}
\newblock
\shownote{arXiv:2309.00267 [cs]}.


\bibitem[Lian et~al\mbox{.}(2023)]%
        {lian_llm-grounded_2023}
\bibfield{author}{\bibinfo{person}{Long Lian}, \bibinfo{person}{Boyi Li}, \bibinfo{person}{Adam Yala}, {and} \bibinfo{person}{Trevor Darrell}.} \bibinfo{year}{2023}\natexlab{}.
\newblock \bibinfo{title}{{LLM}-grounded {Diffusion}: {Enhancing} {Prompt} {Understanding} of {Text}-to-{Image} {Diffusion} {Models} with {Large} {Language} {Models}}.
\newblock
\newblock
\urldef\tempurl%
\url{http://arxiv.org/abs/2305.13655}
\showURL{%
\tempurl}
\newblock
\shownote{arXiv:2305.13655 [cs]}.


\bibitem[Lovato et~al\mbox{.}(2024)]%
        {lovato_foregrounding_2024}
\bibfield{author}{\bibinfo{person}{Juniper Lovato}, \bibinfo{person}{Julia Zimmerman}, \bibinfo{person}{Isabelle Smith}, \bibinfo{person}{Peter Dodds}, {and} \bibinfo{person}{Jennifer Karson}.} \bibinfo{year}{2024}\natexlab{}.
\newblock \bibinfo{title}{Foregrounding {Artist} {Opinions}: {A} {Survey} {Study} on {Transparency}, {Ownership}, and {Fairness} in {AI} {Generative} {Art}}.
\newblock
\newblock
\urldef\tempurl%
\url{https://doi.org/10.48550/arXiv.2401.15497}
\showDOI{\tempurl}
\newblock
\shownote{arXiv:2401.15497}.


\bibitem[Mason(2023)]%
        {pluralism_mason_2023}
\bibfield{author}{\bibinfo{person}{Elinor Mason}.} \bibinfo{year}{2023}\natexlab{}.
\newblock \showarticletitle{Value {Pluralism}}.
\newblock In \bibinfo{booktitle}{\emph{The {Stanford} {Encyclopedia} of {Philosophy}} (\bibinfo{edition}{summer 2023} ed.)}, \bibfield{editor}{\bibinfo{person}{Edward~N. Zalta} {and} \bibinfo{person}{Uri Nodelman}} (Eds.). \bibinfo{publisher}{Metaphysics Research Lab, Stanford University}.
\newblock
\urldef\tempurl%
\url{https://plato.stanford.edu/archives/sum2023/entries/value-pluralism/}
\showURL{%
\tempurl}


\bibitem[Mazzone and Elgammal(2019)]%
        {mazzone_art_2019}
\bibfield{author}{\bibinfo{person}{Marian Mazzone} {and} \bibinfo{person}{Ahmed Elgammal}.} \bibinfo{year}{2019}\natexlab{}.
\newblock \showarticletitle{Art, {Creativity}, and the {Potential} of {Artificial} {Intelligence}}.
\newblock \bibinfo{journal}{\emph{Arts}}  \bibinfo{volume}{8} (\bibinfo{date}{Feb.} \bibinfo{year}{2019}).
\newblock
\urldef\tempurl%
\url{https://doi.org/10.3390/arts8010026}
\showDOI{\tempurl}


\bibitem[McDonald and Fisher(2006)]%
        {art_literacy}
\bibfield{author}{\bibinfo{person}{Nan~Leslie McDonald} {and} \bibinfo{person}{Douglas Fisher}.} \bibinfo{year}{2006}\natexlab{}.
\newblock \bibinfo{booktitle}{\emph{Teaching {Literacy} {Through} the {Arts}}}.
\newblock \bibinfo{publisher}{Guilford Press}.
\newblock
\showISBNx{978-1-59385-280-1}
\newblock
\shownote{Google-Books-ID: lrQukcPC\_zYC}.


\bibitem[Meta(2024)]%
        {MovieGen}
\bibfield{author}{\bibinfo{person}{Meta}.} \bibinfo{year}{2024}\natexlab{}.
\newblock \bibinfo{title}{{Movie} {Gen}: {A} {Cast} of {Media} {Foundation} {Models}}.
\newblock \bibinfo{howpublished}{AI at Meta}.
\newblock
\newblock
\shownote{https://ai.meta.com/static-resource/movie-gen-research-paper}.


\bibitem[Mitchell(1995)]%
        {mitchell_picture_theory_1995}
\bibfield{author}{\bibinfo{person}{W.~J.~T. Mitchell}.} \bibinfo{year}{1995}\natexlab{}.
\newblock \bibinfo{booktitle}{\emph{Picture {Theory}: {Essays} on {Verbal} and {Visual} {Representation}}}.
\newblock \bibinfo{publisher}{University of Chicago Press}, \bibinfo{address}{Chicago, IL}.
\newblock
\showISBNx{978-0-226-53232-5}
\urldef\tempurl%
\url{https://press.uchicago.edu/ucp/books/book/chicago/P/bo3683962.html}
\showURL{%
\tempurl}


\bibitem[Morales-Navarro et~al\mbox{.}(2023)]%
        {morales-navarro_making_nodate}
\bibfield{author}{\bibinfo{person}{Luis Morales-Navarro}, \bibinfo{person}{Yasmin~B Kafai}, \bibinfo{person}{Francisco Castro}, \bibinfo{person}{William Payne}, \bibinfo{person}{Kayla DesPortes}, \bibinfo{person}{Daniella DiPaola}, \bibinfo{person}{Randi Williams}, \bibinfo{person}{Safinah Ali}, \bibinfo{person}{Cynthia Breazeal}, \bibinfo{person}{Clifford Lee}, \bibinfo{person}{Elisabeth Soep}, \bibinfo{person}{YR Media}, \bibinfo{person}{Duri Long}, \bibinfo{person}{Brian Magerko}, \bibinfo{person}{Jaemarie Solyst}, \bibinfo{person}{Amy Ogan}, \bibinfo{person}{Cansu Tatar}, \bibinfo{person}{Shiyan Jiang}, \bibinfo{person}{Jie Chao}, \bibinfo{person}{Carolyn~P Rosé}, {and} \bibinfo{person}{Sepehr Vakil}.} \bibinfo{year}{2023}\natexlab{}.
\newblock \bibinfo{title}{Making {Sense} of {Machine} {Learning}: {Integrating} {Youth}’s {Conceptual}, {Creative}, and {Critical} {Understandings} of {AI}}.
\newblock
\newblock
\showeprint[arxiv]{2305.02840}~[cs.CY]


\bibitem[Moran(2014)]%
        {moran_introduction_2014}
\bibfield{author}{\bibinfo{person}{Seana Moran}.} \bibinfo{year}{2014}\natexlab{}.
\newblock \showarticletitle{Introduction: {The} {Crossroads} of {Creativity} and {Ethics}}.
\newblock In \bibinfo{booktitle}{\emph{The {Ethics} of {Creativity}}}, \bibfield{editor}{\bibinfo{person}{Seana Moran}, \bibinfo{person}{David Cropley}, {and} \bibinfo{person}{James~C. Kaufman}} (Eds.). \bibinfo{publisher}{Palgrave Macmillan UK}, \bibinfo{address}{London}, \bibinfo{pages}{1--22}.
\newblock
\showISBNx{978-1-137-33354-4}
\urldef\tempurl%
\url{https://doi.org/10.1057/9781137333544_1}
\showDOI{\tempurl}


\bibitem[Musschenga(2005)]%
        {musschenga_empirical_2005}
\bibfield{author}{\bibinfo{person}{Albert Musschenga}.} \bibinfo{year}{2005}\natexlab{}.
\newblock \showarticletitle{Empirical {Ethics}, {Context}-{Sensitivity}, and {Contextualism}}.
\newblock \bibinfo{journal}{\emph{The Journal of Medicine and Philosophy}} \bibinfo{volume}{30}, \bibinfo{number}{5} (\bibinfo{date}{Oct.} \bibinfo{year}{2005}), \bibinfo{pages}{467--490}.
\newblock
\showISSN{0360-5310}
\urldef\tempurl%
\url{https://doi.org/10.1080/03605310500253030}
\showDOI{\tempurl}


\bibitem[Ng et~al\mbox{.}(2021)]%
        {ng_conceptualizing_2021}
\bibfield{author}{\bibinfo{person}{Davy Tsz~Kit Ng}, \bibinfo{person}{Jac Ka~Lok Leung}, \bibinfo{person}{Samuel Kai~Wah Chu}, {and} \bibinfo{person}{Maggie~Shen Qiao}.} \bibinfo{year}{2021}\natexlab{}.
\newblock \showarticletitle{Conceptualizing {AI} literacy: {An} exploratory review}.
\newblock \bibinfo{journal}{\emph{Computers and Education: Artificial Intelligence}}  \bibinfo{volume}{2} (\bibinfo{year}{2021}), \bibinfo{pages}{100041}.
\newblock
\showISSN{2666920X}
\urldef\tempurl%
\url{https://doi.org/10.1016/j.caeai.2021.100041}
\showDOI{\tempurl}


\bibitem[Offenhuber(2023)]%
        {offenhuber_autographic_2023}
\bibfield{author}{\bibinfo{person}{Dietmar Offenhuber}.} \bibinfo{year}{2023}\natexlab{}.
\newblock \bibinfo{booktitle}{\emph{Autographic {Design}: {The} {Matter} of {Data} in a {Self}-{Inscribing} {World}}}.
\newblock \bibinfo{publisher}{MIT Press}.
\newblock
\showISBNx{978-0-262-54702-4}
\newblock
\shownote{Google-Books-ID: bpO0EAAAQBAJ}.


\bibitem[OpenAI(2024)]%
        {openai_train_2024}
\bibfield{author}{\bibinfo{person}{OpenAI}.} \bibinfo{year}{2024}\natexlab{}.
\newblock \bibinfo{title}{How your data is used to improve model performance}.
\newblock
\newblock
\urldef\tempurl%
\url{https://help.openai.com/en/articles/5722486-how-your-data-is-used-to-improve-model-performance}
\showURL{%
\tempurl}


\bibitem[Ouyang et~al\mbox{.}(2022)]%
        {ouyang_training_2022}
\bibfield{author}{\bibinfo{person}{Long Ouyang}, \bibinfo{person}{Jeff Wu}, \bibinfo{person}{Xu Jiang}, \bibinfo{person}{Diogo Almeida}, \bibinfo{person}{Carroll~L. Wainwright}, \bibinfo{person}{Pamela Mishkin}, \bibinfo{person}{Chong Zhang}, \bibinfo{person}{Sandhini Agarwal}, \bibinfo{person}{Katarina Slama}, \bibinfo{person}{Alex Ray}, \bibinfo{person}{John Schulman}, \bibinfo{person}{Jacob Hilton}, \bibinfo{person}{Fraser Kelton}, \bibinfo{person}{Luke Miller}, \bibinfo{person}{Maddie Simens}, \bibinfo{person}{Amanda Askell}, \bibinfo{person}{Peter Welinder}, \bibinfo{person}{Paul Christiano}, \bibinfo{person}{Jan Leike}, {and} \bibinfo{person}{Ryan Lowe}.} \bibinfo{year}{2022}\natexlab{}.
\newblock \bibinfo{title}{Training language models to follow instructions with human feedback}.
\newblock
\newblock
\urldef\tempurl%
\url{http://arxiv.org/abs/2203.02155}
\showURL{%
\tempurl}
\newblock
\shownote{arXiv:2203.02155 [cs]}.


\bibitem[Pedreschi et~al\mbox{.}(2025)]%
        {co-evolution}
\bibfield{author}{\bibinfo{person}{Dino Pedreschi}, \bibinfo{person}{Luca Pappalardo}, \bibinfo{person}{Emanuele Ferragina}, \bibinfo{person}{Ricardo Baeza-Yates}, \bibinfo{person}{Albert-László Barabási}, \bibinfo{person}{Frank Dignum}, \bibinfo{person}{Virginia Dignum}, \bibinfo{person}{Tina Eliassi-Rad}, \bibinfo{person}{Fosca Giannotti}, \bibinfo{person}{János Kertész}, \bibinfo{person}{Alistair Knott}, \bibinfo{person}{Yannis Ioannidis}, \bibinfo{person}{Paul Lukowicz}, \bibinfo{person}{Andrea Passarella}, \bibinfo{person}{Alex~Sandy Pentland}, \bibinfo{person}{John Shawe-Taylor}, {and} \bibinfo{person}{Alessandro Vespignani}.} \bibinfo{year}{2025}\natexlab{}.
\newblock \showarticletitle{Human-AI coevolution}.
\newblock \bibinfo{journal}{\emph{Artificial Intelligence}}  \bibinfo{volume}{339} (\bibinfo{year}{2025}), \bibinfo{pages}{104244}.
\newblock
\showISSN{0004-3702}
\urldef\tempurl%
\url{https://doi.org/10.1016/j.artint.2024.104244}
\showDOI{\tempurl}


\bibitem[Peebles and Xie(2023)]%
        {diffusionmodelstransformers}
\bibfield{author}{\bibinfo{person}{William Peebles} {and} \bibinfo{person}{Saining Xie}.} \bibinfo{year}{2023}\natexlab{}.
\newblock \bibinfo{title}{Scalable Diffusion Models with Transformers}.
\newblock
\newblock
\showeprint[arxiv]{2212.09748}~[cs.CV]
\urldef\tempurl%
\url{https://arxiv.org/abs/2212.09748}
\showURL{%
\tempurl}


\bibitem[Qadri et~al\mbox{.}(2023)]%
        {qadri_ais_2023}
\bibfield{author}{\bibinfo{person}{Rida Qadri}, \bibinfo{person}{Renee Shelby}, \bibinfo{person}{Cynthia~L. Bennett}, {and} \bibinfo{person}{Emily Denton}.} \bibinfo{year}{2023}\natexlab{}.
\newblock \showarticletitle{{AI}’s {Regimes} of {Representation}: {A} {Community}-centered {Study} of {Text}-to-{Image} {Models} in {South} {Asia}}. In \bibinfo{booktitle}{\emph{2023 {ACM} {Conference} on {Fairness}, {Accountability}, and {Transparency}}}. \bibinfo{publisher}{ACM}, \bibinfo{address}{Chicago IL USA}, \bibinfo{pages}{506--517}.
\newblock
\showISBNx{9798400701924}
\urldef\tempurl%
\url{https://doi.org/10.1145/3593013.3594016}
\showDOI{\tempurl}


\bibitem[Raghavan(2024)]%
        {gemini}
\bibfield{author}{\bibinfo{person}{Prabhakar Raghavan}.} \bibinfo{year}{2024}\natexlab{}.
\newblock \bibinfo{title}{Gemini image generation got it wrong. {We}'ll do better.}
\newblock
\newblock
\urldef\tempurl%
\url{https://blog.google/products/gemini/gemini-image-generation-issue/}
\showURL{%
\tempurl}


\bibitem[Ranjita and Nushi(2023)]%
        {naik_social_2023}
\bibfield{author}{\bibinfo{person}{Ranjita} {and} \bibinfo{person}{Besmira Nushi}.} \bibinfo{year}{2023}\natexlab{}.
\newblock \showarticletitle{Social {Biases} through the {Text}-to-{Image} {Generation} {Lens}}. In \bibinfo{booktitle}{\emph{Proceedings of the 2023 {AAAI}/{ACM} {Conference} on {AI}, {Ethics}, and {Society}}}. \bibinfo{publisher}{ACM}, \bibinfo{address}{Montr{\textbackslash}'\{e\}al QC Canada}, \bibinfo{pages}{786--808}.
\newblock
\showISBNx{9798400702310}
\urldef\tempurl%
\url{https://doi.org/10.1145/3600211.3604711}
\showDOI{\tempurl}


\bibitem[Richards and Hemphill({[n.\,d.]})]%
        {richards_practical_nodate}
\bibfield{author}{\bibinfo{person}{K.~Andrew~R. Richards} {and} \bibinfo{person}{Michael~A. Hemphill}.} \bibinfo{year}{[n.\,d.]}\natexlab{}.
\newblock \showarticletitle{A Practical Guide to Collaborative Qualitative Data Analysis}.
\newblock  \bibinfo{volume}{37}, \bibinfo{number}{2} (\bibinfo{year}{[n.\,d.]}), \bibinfo{pages}{225--231}.
\newblock
\showISSN{1543-2769, 0273-5024}
\urldef\tempurl%
\url{https://doi.org/10.1123/jtpe.2017-0084}
\showDOI{\tempurl}
\newblock
\shownote{Publisher: Human Kinetics Section: Journal of Teaching in Physical Education}.


\bibitem[Rojas et~al\mbox{.}(2022)]%
        {rojas_dollar_nodate}
\bibfield{author}{\bibinfo{person}{William A~Gaviria Rojas}, \bibinfo{person}{Sudnya Diamos}, \bibinfo{person}{Keertan~Ranjan Kini}, \bibinfo{person}{David Kanter}, \bibinfo{person}{Vijay~Janapa Reddi}, {and} \bibinfo{person}{Cody Coleman}.} \bibinfo{year}{2022}\natexlab{}.
\newblock \showarticletitle{The {Dollar} {Street} {Dataset}: {Images} {Representing} the {Geographic} and {Socioeconomic} {Diversity} of the {World}}.
\newblock \bibinfo{journal}{\emph{Thirty-sixth Conference on Neural Information Processing Systems Datasets and Benchmarks Track}} (\bibinfo{year}{2022}).
\newblock


\bibitem[Rusu(2017)]%
        {rusu_art_empathy}
\bibfield{author}{\bibinfo{person}{Marinela Rusu}.} \bibinfo{year}{2017}\natexlab{}.
\newblock \showarticletitle{2. {Empathy} and {Communication} through {Art}}.
\newblock \bibinfo{journal}{\emph{Review of Artistic Education}}  \bibinfo{volume}{14} (\bibinfo{date}{March} \bibinfo{year}{2017}).
\newblock
\urldef\tempurl%
\url{https://doi.org/10.1515/rae-2017-0018}
\showDOI{\tempurl}


\bibitem[Saldana(2021)]%
        {saldana_coding_2021}
\bibfield{author}{\bibinfo{person}{Johnny Saldana}.} \bibinfo{year}{2021}\natexlab{}.
\newblock \showarticletitle{The Coding Manual for Qualitative Researchers}.
\newblock  (\bibinfo{year}{2021}), \bibinfo{pages}{1--440}.
\newblock
\urldef\tempurl%
\url{https://www.torrossa.com/en/resources/an/5018667}
\showURL{%
\tempurl}
\newblock
\shownote{Publisher: {SAGE} Publications Ltd}.


\bibitem[Scott(2009)]%
        {scott_surrogacy_2009}
\bibfield{author}{\bibinfo{person}{Elizabeth~S Scott}.} \bibinfo{year}{2009}\natexlab{}.
\newblock \showarticletitle{Surrogacy and the {Politics} of {Commodification}}.
\newblock \bibinfo{journal}{\emph{LAW AND CONTEMPORARY PROBLEMS}}  \bibinfo{volume}{72} (\bibinfo{year}{2009}).
\newblock


\bibitem[Selbst et~al\mbox{.}(2019)]%
        {selbst_fairness_2019}
\bibfield{author}{\bibinfo{person}{Andrew~D. Selbst}, \bibinfo{person}{Danah Boyd}, \bibinfo{person}{Sorelle~A. Friedler}, \bibinfo{person}{Suresh Venkatasubramanian}, {and} \bibinfo{person}{Janet Vertesi}.} \bibinfo{year}{2019}\natexlab{}.
\newblock \showarticletitle{Fairness and {Abstraction} in {Sociotechnical} {Systems}}. In \bibinfo{booktitle}{\emph{Proceedings of the {Conference} on {Fairness}, {Accountability}, and {Transparency}}}. \bibinfo{publisher}{ACM}, \bibinfo{address}{Atlanta GA USA}, \bibinfo{pages}{59--68}.
\newblock
\showISBNx{978-1-4503-6125-5}
\urldef\tempurl%
\url{https://doi.org/10.1145/3287560.3287598}
\showDOI{\tempurl}


\bibitem[Seymour et~al\mbox{.}(2022)]%
        {seymour_respect_2022}
\bibfield{author}{\bibinfo{person}{William Seymour}, \bibinfo{person}{Max Van~Kleek}, \bibinfo{person}{Reuben Binns}, {and} \bibinfo{person}{Dave Murray-Rust}.} \bibinfo{year}{2022}\natexlab{}.
\newblock \showarticletitle{Respect as a {Lens} for the {Design} of {AI} {Systems}}. In \bibinfo{booktitle}{\emph{Proceedings of the 2022 {AAAI}/{ACM} {Conference} on {AI}, {Ethics}, and {Society}}}. \bibinfo{publisher}{ACM}, \bibinfo{address}{Oxford United Kingdom}, \bibinfo{pages}{641--652}.
\newblock
\showISBNx{978-1-4503-9247-1}
\urldef\tempurl%
\url{https://doi.org/10.1145/3514094.3534186}
\showDOI{\tempurl}


\bibitem[Shenton({[n.\,d.]})]%
        {shenton_strategies_2004}
\bibfield{author}{\bibinfo{person}{Andrew~K. Shenton}.} \bibinfo{year}{[n.\,d.]}\natexlab{}.
\newblock \showarticletitle{Strategies for ensuring trustworthiness in qualitative research projects}.
\newblock  \bibinfo{volume}{22}, \bibinfo{number}{2} (\bibinfo{year}{[n.\,d.]}), \bibinfo{pages}{63--75}.
\newblock
\showISSN{0167-8329}
\urldef\tempurl%
\url{https://doi.org/10.3233/EFI-2004-22201}
\showDOI{\tempurl}
\newblock
\shownote{Publisher: {IOS} Press}.


\bibitem[Shumailov et~al\mbox{.}(2024)]%
        {shumailov_ai_2024}
\bibfield{author}{\bibinfo{person}{Ilia Shumailov}, \bibinfo{person}{Zakhar Shumaylov}, \bibinfo{person}{Yiren Zhao}, \bibinfo{person}{Nicolas Papernot}, \bibinfo{person}{Ross Anderson}, {and} \bibinfo{person}{Yarin Gal}.} \bibinfo{year}{2024}\natexlab{}.
\newblock \showarticletitle{{AI} models collapse when trained on recursively generated data}.
\newblock \bibinfo{journal}{\emph{Nature}} \bibinfo{volume}{631}, \bibinfo{number}{8022} (\bibinfo{date}{July} \bibinfo{year}{2024}), \bibinfo{pages}{755--759}.
\newblock
\showISSN{1476-4687}
\urldef\tempurl%
\url{https://doi.org/10.1038/s41586-024-07566-y}
\showDOI{\tempurl}


\bibitem[Sinnott-Armstrong(2023)]%
        {consequentialism_sinnott-armstrong_2023}
\bibfield{author}{\bibinfo{person}{Walter Sinnott-Armstrong}.} \bibinfo{year}{2023}\natexlab{}.
\newblock \showarticletitle{Consequentialism}.
\newblock In \bibinfo{booktitle}{\emph{The {Stanford} {Encyclopedia} of {Philosophy}} (\bibinfo{edition}{winter 2023} ed.)}, \bibfield{editor}{\bibinfo{person}{Edward~N. Zalta} {and} \bibinfo{person}{Uri Nodelman}} (Eds.). \bibinfo{publisher}{Metaphysics Research Lab, Stanford University}.
\newblock
\urldef\tempurl%
\url{https://plato.stanford.edu/archives/win2023/entries/consequentialism/}
\showURL{%
\tempurl}


\bibitem[Sluis and Molesworth(2023)]%
        {whitney_critical_2023}
\bibfield{author}{\bibinfo{person}{Katrina Sluis} {and} \bibinfo{person}{Erica Molesworth}.} \bibinfo{year}{2023}\natexlab{}.
\newblock \bibinfo{title}{Critical {AI} in the {Art} {Museum}}.
\newblock
\newblock
\urldef\tempurl%
\url{https://criticalai.art/}
\showURL{%
\tempurl}


\bibitem[Sorensen et~al\mbox{.}(2023)]%
        {sorensen_value_2023}
\bibfield{author}{\bibinfo{person}{Taylor Sorensen}, \bibinfo{person}{Liwei Jiang}, \bibinfo{person}{Jena Hwang}, \bibinfo{person}{Sydney Levine}, \bibinfo{person}{Valentina Pyatkin}, \bibinfo{person}{Peter West}, \bibinfo{person}{Nouha Dziri}, \bibinfo{person}{Ximing Lu}, \bibinfo{person}{Kavel Rao}, \bibinfo{person}{Chandra Bhagavatula}, \bibinfo{person}{Maarten Sap}, \bibinfo{person}{John Tasioulas}, {and} \bibinfo{person}{Yejin Choi}.} \bibinfo{year}{2023}\natexlab{}.
\newblock \bibinfo{title}{Value {Kaleidoscope}: {Engaging} {AI} with {Pluralistic} {Human} {Values}, {Rights}, and {Duties}}.
\newblock
\newblock
\urldef\tempurl%
\url{http://arxiv.org/abs/2309.00779}
\showURL{%
\tempurl}
\newblock
\shownote{arXiv:2309.00779 [cs]}.


\bibitem[Srinivasan(2024)]%
        {srinivasan_see_2024}
\bibfield{author}{\bibinfo{person}{Ramya Srinivasan}.} \bibinfo{year}{2024}\natexlab{}.
\newblock \showarticletitle{To {See} or {Not} to {See}: {Understanding} the {Tensions} of {Algorithmic} {Curation} for {Visual} {Arts}}. In \bibinfo{booktitle}{\emph{The 2024 {ACM} {Conference} on {Fairness}, {Accountability}, and {Transparency}}}. \bibinfo{publisher}{ACM}, \bibinfo{address}{Rio de Janeiro Brazil}, \bibinfo{pages}{444--455}.
\newblock
\showISBNx{9798400704505}
\urldef\tempurl%
\url{https://doi.org/10.1145/3630106.3658917}
\showDOI{\tempurl}


\bibitem[Srinivasan and Parikh(2022)]%
        {srinivasan_building_2021}
\bibfield{author}{\bibinfo{person}{Ramya Srinivasan} {and} \bibinfo{person}{Devi Parikh}.} \bibinfo{year}{2022}\natexlab{}.
\newblock \bibinfo{title}{Building {Bridges}: {Generative} {Artworks} to {Explore} {AI} {Ethics}}.
\newblock
\newblock
\urldef\tempurl%
\url{http://arxiv.org/abs/2106.13901}
\showURL{%
\tempurl}
\newblock
\shownote{2022 CVPR Workshop on Ethical Considerations in Creative Applications of Computer Vision}.


\bibitem[Srinivasan and Uchino(2021)]%
        {srinivasan_role_2021}
\bibfield{author}{\bibinfo{person}{Ramya Srinivasan} {and} \bibinfo{person}{Kanji Uchino}.} \bibinfo{year}{2021}\natexlab{}.
\newblock \bibinfo{title}{The {Role} of {Arts} in {Shaping} {AI} {Ethics}}.
\newblock
\newblock
\newblock
\shownote{2021 AAAI Workshop on Reframing Diversity in AI: Representation, Power, and Inclusion}.


\bibitem[Srinivasan and Uchino(2023)]%
        {srinivasan_patent}
\bibfield{author}{\bibinfo{person}{Ramya Srinivasan} {and} \bibinfo{person}{Kanji Uchino}.} \bibinfo{year}{U.S. Patent 20230186535A1, Jun. 2023}\natexlab{}.
\newblock \bibinfo{title}{{Image} {Generation} based on {Ethical} {Viewpoints}}.
\newblock
\newblock


\bibitem[Sterman et~al\mbox{.}(2023)]%
        {sterman_kaleidoscope_2023}
\bibfield{author}{\bibinfo{person}{Sarah Sterman}, \bibinfo{person}{Molly~Jane Nicholas}, \bibinfo{person}{Janaki Vivrekar}, \bibinfo{person}{Jessica~R Mindel}, {and} \bibinfo{person}{Eric Paulos}.} \bibinfo{year}{2023}\natexlab{}.
\newblock \showarticletitle{Kaleidoscope: {A} {Reflective} {Documentation} {Tool} for a {User} {Interface} {Design} {Course}}. In \bibinfo{booktitle}{\emph{Proceedings of the 2023 {CHI} {Conference} on {Human} {Factors} in {Computing} {Systems}}}. \bibinfo{publisher}{ACM}, \bibinfo{address}{Hamburg Germany}, \bibinfo{pages}{1--19}.
\newblock
\showISBNx{978-1-4503-9421-5}
\urldef\tempurl%
\url{https://doi.org/10.1145/3544548.3581255}
\showDOI{\tempurl}


\bibitem[Tendulkar et~al\mbox{.}(2019)]%
        {tendulkar_trick_2019}
\bibfield{author}{\bibinfo{person}{Purva Tendulkar}, \bibinfo{person}{Kalpesh Krishna}, \bibinfo{person}{Ramprasaath~R. Selvaraju}, {and} \bibinfo{person}{Devi Parikh}.} \bibinfo{year}{2019}\natexlab{}.
\newblock \showarticletitle{Trick or {TReAT}: {Thematic} {Reinforcement} for {Artistic} {Typography}}. In \bibinfo{booktitle}{\emph{Proceedings of the 2019 {International} {Conference} on {Computational} {Creativity}}}. \bibinfo{publisher}{ICCC}.
\newblock


\bibitem[Thiroux and Krasemann(2014)]%
        {theory_and_practice}
\bibfield{author}{\bibinfo{person}{Jacques~P. Thiroux} {and} \bibinfo{person}{Keith~W. Krasemann}.} \bibinfo{year}{2014}\natexlab{}.
\newblock \bibinfo{booktitle}{\emph{Ethics: {Theory} and {Practice}} (\bibinfo{edition}{11th edition} ed.)}.
\newblock \bibinfo{publisher}{Pearson}, \bibinfo{address}{Boston}.
\newblock
\showISBNx{978-0-13-380405-8}


\bibitem[Timmons(2012)]%
        {timmons_moral_2012}
\bibfield{author}{\bibinfo{person}{Mark Timmons}.} \bibinfo{year}{2012}\natexlab{}.
\newblock \bibinfo{booktitle}{\emph{Moral {Theory}: {An} {Introduction}}}.
\newblock \bibinfo{publisher}{Rowman \& Littlefield Publishers}.
\newblock
\showISBNx{978-0-7425-6493-0}
\newblock
\shownote{Google-Books-ID: qWGp1iK9IlAC}.


\bibitem[Ukoh et~al\mbox{.}(2022)]%
        {physics}
\bibfield{author}{\bibinfo{person}{Edidiong~Enyeneokpon Ukoh}, \bibinfo{person}{Jude Nicholas}, \bibinfo{person}{Edidiong~Enyeneokpon Ukoh}, {and} \bibinfo{person}{Jude Nicholas}.} \bibinfo{year}{2022}\natexlab{}.
\newblock \showarticletitle{AI Adoption for Teaching and Learning of Physics}.
\newblock  (\bibinfo{year}{2022}).
\newblock
\urldef\tempurl%
\url{https://doi.org/10.20533/IJI.1742.4712.2022.0222}
\showDOI{\tempurl}


\bibitem[Varshney(2024)]%
        {varshney_decolonial_2024}
\bibfield{author}{\bibinfo{person}{Kush~R. Varshney}.} \bibinfo{year}{2024}\natexlab{}.
\newblock \bibinfo{title}{Decolonial {AI} {Alignment}: {Openness}, {Viśe}{\textbackslash}d\{s\}a-{Dharma}, and {Including} {Excluded} {Knowledges}}.
\newblock
\newblock
\urldef\tempurl%
\url{http://arxiv.org/abs/2309.05030}
\showURL{%
\tempurl}
\newblock
\shownote{arXiv:2309.05030}.


\bibitem[Vladova et~al\mbox{.}(2024)]%
        {vladova_why_2024}
\bibfield{author}{\bibinfo{person}{Gergana Vladova}, \bibinfo{person}{Jennifer Haase}, {and} \bibinfo{person}{Sascha Friesike}.} \bibinfo{year}{2024}\natexlab{}.
\newblock \showarticletitle{Why, with whom, and how to conduct interdisciplinary research? {A} review from a researcher’s perspective}.
\newblock \bibinfo{journal}{\emph{Science and Public Policy}} \bibinfo{volume}{52}, \bibinfo{number}{2} (\bibinfo{date}{Nov.} \bibinfo{year}{2024}), \bibinfo{pages}{165--180}.
\newblock
\showISSN{0302-3427}
\urldef\tempurl%
\url{https://doi.org/10.1093/scipol/scae070}
\showDOI{\tempurl}
\newblock
\shownote{\_eprint: https://academic.oup.com/spp/article-pdf/52/2/165/60797068/scae070.pdf}.


\bibitem[Walker et~al\mbox{.}(2023)]%
        {walker_ai_2023}
\bibfield{author}{\bibinfo{person}{Johanna Walker}, \bibinfo{person}{Gefion Thuermer}, \bibinfo{person}{Julian Vicens}, {and} \bibinfo{person}{Elena Simperl}.} \bibinfo{year}{2023}\natexlab{}.
\newblock \showarticletitle{{AI} {Art} and {Misinformation}: {Approaches} and {Strategies} for {Media} {Literacy} and {Fact} {Checking}}. In \bibinfo{booktitle}{\emph{Proceedings of the 2023 {AAAI}/{ACM} {Conference} on {AI}, {Ethics}, and {Society}}}. \bibinfo{publisher}{ACM}, \bibinfo{address}{Montr{\textbackslash}'\{e\}al QC Canada}, \bibinfo{pages}{26--37}.
\newblock
\showISBNx{9798400702310}
\urldef\tempurl%
\url{https://doi.org/10.1145/3600211.3604715}
\showDOI{\tempurl}


\bibitem[Wallach and Vallor(2020)]%
        {wallach_moral_2020}
\bibfield{author}{\bibinfo{person}{Wendell Wallach} {and} \bibinfo{person}{Shannon Vallor}.} \bibinfo{year}{2020}\natexlab{}.
\newblock \showarticletitle{Moral {Machines}: {From} {Value} {Alignment} to {Embodied} {Virtue}}.
\newblock In \bibinfo{booktitle}{\emph{Ethics of {Artificial} {Intelligence}}}, \bibfield{editor}{\bibinfo{person}{S.~Matthew Liao}} (Ed.). \bibinfo{publisher}{Oxford University Press}, \bibinfo{pages}{0}.
\newblock
\showISBNx{978-0-19-090503-3}
\urldef\tempurl%
\url{https://doi.org/10.1093/oso/9780190905033.003.0014}
\showDOI{\tempurl}


\bibitem[Wang et~al\mbox{.}(2024)]%
        {wang_transform_2024}
\bibfield{author}{\bibinfo{person}{Jiajun Wang}, \bibinfo{person}{Morteza Ghahremani}, \bibinfo{person}{Yitong Li}, \bibinfo{person}{Björn Ommer}, {and} \bibinfo{person}{Christian Wachinger}.} \bibinfo{year}{2024}\natexlab{}.
\newblock \bibinfo{title}{Stable-Pose: Leveraging Transformers for Pose-Guided Text-to-Image Generation}.
\newblock
\newblock
\showeprint[arxiv]{2406.02485}~[cs.CV]
\urldef\tempurl%
\url{https://arxiv.org/abs/2406.02485}
\showURL{%
\tempurl}


\bibitem[Wang and Yin(2023)]%
        {wang_watch_2023}
\bibfield{author}{\bibinfo{person}{Xinru Wang} {and} \bibinfo{person}{Ming Yin}.} \bibinfo{year}{2023}\natexlab{}.
\newblock \showarticletitle{Watch {Out} for {Updates}: {Understanding} the {Effects} of {Model} {Explanation} {Updates} in {AI}-{Assisted} {Decision} {Making}}. In \bibinfo{booktitle}{\emph{Proceedings of the 2023 {CHI} {Conference} on {Human} {Factors} in {Computing} {Systems}}} \emph{(\bibinfo{series}{{CHI} '23})}. \bibinfo{publisher}{Association for Computing Machinery}, \bibinfo{address}{New York, NY, USA}, \bibinfo{pages}{1--19}.
\newblock
\showISBNx{978-1-4503-9421-5}
\urldef\tempurl%
\url{https://doi.org/10.1145/3544548.3581366}
\showDOI{\tempurl}


\bibitem[West et~al\mbox{.}(2023)]%
        {west2023generative_understand}
\bibfield{author}{\bibinfo{person}{Peter West}, \bibinfo{person}{Ximing Lu}, \bibinfo{person}{Nouha Dziri}, \bibinfo{person}{Faeze Brahman}, \bibinfo{person}{Linjie Li}, \bibinfo{person}{Jena~D. Hwang}, \bibinfo{person}{Liwei Jiang}, \bibinfo{person}{Jillian Fisher}, \bibinfo{person}{Abhilasha Ravichander}, \bibinfo{person}{Khyathi Chandu}, \bibinfo{person}{Benjamin Newman}, \bibinfo{person}{Pang~Wei Koh}, \bibinfo{person}{Allyson Ettinger}, {and} \bibinfo{person}{Yejin Choi}.} \bibinfo{year}{2023}\natexlab{}.
\newblock \bibinfo{title}{The Generative AI Paradox: "What It Can Create, It May Not Understand"}.
\newblock
\newblock
\showeprint[arxiv]{2311.00059}~[cs.AI]


\bibitem[Wolfe and Mitra(2024)]%
        {wolfe_implications_2024}
\bibfield{author}{\bibinfo{person}{Robert Wolfe} {and} \bibinfo{person}{Tanushree Mitra}.} \bibinfo{year}{2024}\natexlab{}.
\newblock \showarticletitle{The {Implications} of {Open} {Generative} {Models} in {Human}-{Centered} {Data} {Science} {Work}: {A} {Case} {Study} with {Fact}-{Checking} {Organizations}}. In \bibinfo{booktitle}{\emph{Proceedings of the {AAAI}/{ACM} {Conference} on {AI}, {Ethics}, and {Society}}}, Vol.~\bibinfo{volume}{7}. \bibinfo{pages}{1595--1607}.
\newblock
\urldef\tempurl%
\url{https://ojs.aaai.org/index.php/AIES/article/view/31750}
\showURL{%
\tempurl}


\bibitem[Wong(2024)]%
        {wong_ai-generated_2024}
\bibfield{author}{\bibinfo{person}{Carissa Wong}.} \bibinfo{year}{2024}\natexlab{}.
\newblock \showarticletitle{{AI}-generated images and video are here: how could they shape research?}
\newblock \bibinfo{journal}{\emph{Nature}} (\bibinfo{date}{March} \bibinfo{year}{2024}).
\newblock
\urldef\tempurl%
\url{https://doi.org/10.1038/d41586-024-00659-8}
\showDOI{\tempurl}
\newblock
\shownote{Bandiera\_abtest: a Cg\_type: News Explainer Publisher: Nature Publishing Group Subject\_term: Scientific community, Software, Machine learning}.


\bibitem[Wu et~al\mbox{.}(2023)]%
        {wu_honor_2023}
\bibfield{author}{\bibinfo{person}{Stephen Tze-Inn Wu}, \bibinfo{person}{Daniel Demetriou}, {and} \bibinfo{person}{Rudwan~Ali Husain}.} \bibinfo{year}{2023}\natexlab{}.
\newblock \showarticletitle{Honor {Ethics}: {The} {Challenge} of {Globalizing} {Value} {Alignment} in {AI}}. In \bibinfo{booktitle}{\emph{2023 {ACM} {Conference} on {Fairness}, {Accountability}, and {Transparency}}}. \bibinfo{publisher}{ACM}, \bibinfo{address}{Chicago IL USA}, \bibinfo{pages}{593--602}.
\newblock
\showISBNx{9798400701924}
\urldef\tempurl%
\url{https://doi.org/10.1145/3593013.3594026}
\showDOI{\tempurl}


\bibitem[Yao et~al\mbox{.}(2023)]%
        {yao_unlearning_2023}
\bibfield{author}{\bibinfo{person}{Yuanshun Yao}, \bibinfo{person}{Xiaojun Xu}, {and} \bibinfo{person}{Yang Liu}.} \bibinfo{year}{2023}\natexlab{}.
\newblock \bibinfo{title}{Large {Language} {Model} {Unlearning}}.
\newblock
\newblock
\urldef\tempurl%
\url{https://doi.org/10.48550/arXiv.2310.10683}
\showDOI{\tempurl}
\newblock
\shownote{arXiv:2310.10683 [cs]}.


\bibitem[Zhang et~al\mbox{.}(2023)]%
        {T2I_survey}
\bibfield{author}{\bibinfo{person}{Chenshuang Zhang}, \bibinfo{person}{Chaoning Zhang}, \bibinfo{person}{Mengchun Zhang}, {and} \bibinfo{person}{In~So Kweon}.} \bibinfo{year}{2023}\natexlab{}.
\newblock \bibinfo{title}{Text-to-image {Diffusion} {Models} in {Generative} {AI}: {A} {Survey}}.
\newblock
\newblock
\urldef\tempurl%
\url{http://arxiv.org/abs/2303.07909}
\showURL{%
\tempurl}
\newblock
\shownote{arXiv:2303.07909 [cs]}.


\bibitem[Zhang et~al\mbox{.}(2017)]%
        {zhang_ornamental_2017}
\bibfield{author}{\bibinfo{person}{Junsong Zhang}, \bibinfo{person}{Yu Wang}, \bibinfo{person}{Weiyi Xiao}, {and} \bibinfo{person}{Zhenshan Luo}.} \bibinfo{year}{2017}\natexlab{}.
\newblock \showarticletitle{Synthesizing {Ornamental} {Typefaces}}.
\newblock \bibinfo{journal}{\emph{Computer Graphics Forum}} \bibinfo{volume}{36}, \bibinfo{number}{1} (\bibinfo{year}{2017}), \bibinfo{pages}{64--75}.
\newblock
\showISSN{1467-8659}
\urldef\tempurl%
\url{https://doi.org/10.1111/cgf.12785}
\showDOI{\tempurl}
\newblock
\shownote{\_eprint: https://onlinelibrary.wiley.com/doi/pdf/10.1111/cgf.12785}.


\end{thebibliography}

\newpage
\appendix 
\section{FORMATIVE STUDY INTERVIEW SCRIPT}

\title{Formative Study Interview Script} 

\subsection*{1. Introduction}

Hello [Insert name]!

Pleasure to meet you. Thank you for joining me in conversation today!

My name is [name], and I am [occupation] in [Institution]. I look forward to our conversation today!

Before we begin, I will detail the study we’re to embark on. This research is at the intersection of philosophy, art, and AI and seeks to understand Ethical Theories and their kaleidoscopic nature within Generative AI models. We will use a sect of Generative AI models known as text-to-image models (T2I) to visualize ethical theories in their visual syn- (synthesis/synthetic) generation.

Ethical theories, as we see, are subject to change and this is where we establish our first kaleidoscope. The shifting nature of what we define and practice as those theories themselves. Likewise, Generative AI models are not immune to that nature. They are changing through user input and interaction, model training, where the images you see now may not be the images you see later. They accord to conceptions of the world in accordance with the world at play.

I also have a kaleidoscope here with me to visualize our metaphor. I will pass it to you so you may look into it and ask about its resemblance to the study.

Feel free to answer only the questions you’re comfortable with. If there’s a question that you don’t want to discuss, please let me know. We will also audio record this interview for thematic analysis and I will contact you upon this study with images of the various theories to solicit your input on the visual synthesis of the generated imagery.

If all is well, I’m happy to begin whenever you are! 

\subsection*{2. Interviewee}

I’d like to begin by allowing you to introduce yourself and the academic discipline you come from. 

[Detail answer]

Awesome! Thank you for that. It’s a pleasure to get to know you!

\subsection*{3.	Ethical Theories}

Let’s begin with Ethical Theories. 

\begin{itemize}
    \item To your understanding, what are ethical theories as a synthesis, and you can dive back into what ethics are in the first place? This is what you conceptualize as theories prior to categories of what may lie underneath.
    \item Do you construct a difference to ethical theories in the first place? Is there a difference to ethical theories or a common place that is shared for guidance to process and decisions?
\end{itemize}

I see. So, to proceed:

\begin{itemize}
    \item What Ethical Theories do you identify?
    \item Are you familiar with lesser-known theories and if so please detail your knowledge and awareness of them?
    \item Are there Ethical Theories that are more prominent in your discipline (for example in the law) that you’re more familiar with?
    \item Have you identified differences in theories or has it been a way in which you were taught and adhered to?
\end{itemize}

Okay, awesome. Thank you for sharing your foundations for the theories. It’s really important to look at it from multiple perspectives. So, if I may continue:

Please list some Ethical Theories that you are aware of and their corresponding definitions. 

\textit{[Depending on the breadth of theories from above, we may list some and then iteratively proceed with definitions]}

\begin{itemize}
    \item Utilitarian, \textit{the emphasis is on maximizing the well-being of all stakeholders.}
    \item Deontological, \textit{the emphasis is on following the laws and regulations.}
    \item Virtue, \textit{the emphasis is on moral values.}
    \item Subjectivism, \textit{the emphasis is on individual beliefs, experiences, and opinions.}
    \item Care, \textit{the emphasis is on interpersonal aspects such as caring, interdependence, and the ethical requirements of particular relationships.}
    \item Situational, \textit{takes into account the particular context of an act, rather than judging it according to absolute moral standards or normative rules.}
    \item Consequential, \textit{focuses on the consequences of an act.}
\end{itemize}

\subsection*{4.	Examples of Ethical Theories}

Thank you so much for sharing your insight. There are very important cases in which what we identify as ethics may have, or not have been divided as such. The distinctions make us aware. 

Now, to may proceed. We’re looking to define these definitions of ethics because we see ethics in two ways. One in definition, as to how we conceptualize them and in practice, how we understand them and govern them in our lives. In that pursuit, what we seek to do next is refine ethics in definition and in practice. 

From those above, we will now elicit examples of what could be praxis to ethical theories. Feel free to take your time and we will go in order of the theories you have identified as well:

\textit{[Mentioned theories and ethics that have been identified by the participant. If all:]}

\begin{itemize}
    \item Utilitarian, e.g. human rights
    \item Deontological, e.g. thou shall not kill
    \item Virtue, e.g. compassion, grace
    \item Subjectivism, e.g. experience
    \item Care, e.g. thinking about others
    \item Situational, e.g. momentary
    \item Consequential, e.g. to do harm
\end{itemize}

Awesome. Thank you for sharing those understandings and your insight! I hope you found this study insightful. Before I close, adding perspectives of richness:

\begin{itemize}
    \item Is there anything you would like to detail at this point, as well as any references for us to proceed?
\end{itemize}

Well, I truly appreciate your taking your time and energy within our study. Please don’t hesitate to reach out if you have any questions or follow-up that might have seeded any thoughts. 

Thank you again for participating in our study and I wish you a wonderful day!

\textit{End of Interview}


\end{document}